\documentclass{jpsj3}
\usepackage{amsmath}
\usepackage{graphicx}
\usepackage{dcolumn}
\usepackage{bm}

\title{Spin-Density-Wave Gap with Dirac Nodes and Two-Magnon Raman Scattering in 
BaFe$_2$As$_2$}

\author{Shunji Sugai\thanks{E-mail address: sugai.shunji@h.mbox.nagoya-u.ac.jp, ssugai@pi.ac.ae}$^{1,2,3}$, 
Yuki Mizuno$^{1,3}$, Ryoutarou Watanabe$^{3,4}$, Takahiko Kawaguchi$^{3,4}$, 
Koshi Takenaka$^{3,4}$, Hiroshi Ikuta$^{3,4}$, Yasumasa Takayanagi$^1$, 
Naoki Hayamizu$^1$, and Yasuhiro Sone$^1$} 

\inst{\address{$^1$Department of Physics, Faculty of Science, 
Nagoya University, Furo-cho, Chikusa-ku, Nagoya 464-8602, Japan \\
$^2$Department of Physics, Arts and Science, Petroleum Institute, 
P.O.Box 2533, Abu Dhabi, UAE \\
$^3$TRIP, Japan Science and Technology Agency (JST), Chiyoda, 
Tokyo 102-0075, Japan \\
$^4$Department of Crystalline Materials Science, Nagoya University, 
Furo-cho, Chikusa-ku, Nagoya 464-8603, Japan
}}

\abst{
Raman selection rules for electronic and magnetic excitations in BaFe$_2$As$_2$ 
were theoretically investigated and applied them to the separate detection of 
the nodal and anti-nodal gap excitations at the spin density wave (SDW) transition 
and the separate detection of the nearest and the next nearest neighbor exchange 
interaction energies.  
Raman spectra are composed of magnetic excitations with gradually decreasing 
intensity toward far above the SDW transition temperature 
($T_{\rm SDW}$) and electronic excitations induced by the Brillouin zone 
folding below $T_{\rm SDW}$.  
The SDW gap has Dirac nodes, because many orbitals participate in the electronic 
states near the Fermi energy.  
Using a two-orbital band model the electronic excitations near the Dirac node and 
the anti-node are found to have different symmetries.  
Applying the symmetry difference to Raman scattering the nodal and anti-nodal 
electronic excitations are separately obtained.  
The low-energy spectra from the anti-nodal region have critical fluctuation 
just above $T_{\rm SDW}$ and change into the gap 
structure by the first order transition at $T_{\rm SDW}$, while those from 
the nodal region gradually change into the SDW state.  
Magnetic excitations are observed as a very broad peak in all polarization 
configurations.  
The selection rule for two-magnon scattering from the stripe spin structure was 
obtained.  
Applying it to the two-magnon Raman spectra it is found that the magnetic 
exchange interaction energies are not presented by the short-range superexchange 
model, but the second derivative of the total energy of the stripe spin structure 
with respect to the moment directions.  
The selection rule and the peak energy are expressed by the two-magnon 
scattering process in an insulator, but the large spectral weight above twice the 
maximum spin wave energy is difficult to explain by the decayed spin wave.  
It may be explained by the electronic scattering of itinerant carriers with 
the magnetic self-energy in the localized spin picture or the 
particle-hole excitation model in the itinerant spin picture.  
The magnetic scattering spectra are compared to the insulating and metallic 
cuprate superconductors whose spins are believed to be localized.  
}

\kword{BaFe$_2$As$_2$, SDW gap, Dirac node, Two-magnon Raman scattering}

\begin{document}
\maketitle

\section{Introduction}

Two-dimensional iron pnictides form a new family of superconductors with the 
transition temperature up to $T_{\rm c}=55$ K \cite{Kamihara,Ren}.  
The BaFe$_2$As$_2$ family undergoes superconducting state when the spin 
density wave (SDW) is suppressed by substituting the 
element \cite{Chen,Rotter,Rotter2} or pressurizing \cite{Torikachvili}.  
Quasi-static magnetic measurement shows the disappearance of the spin order and 
magnetic excitations above the spin density wave transition temperature 
$T_{\rm SDW}$ or in the metallic phase.  
However, the lost spin order is only perpendicular to the two-dimensional 
layer \cite{Thio} and the high-energy magnetic excitations in the layer 
clearly remains above $T_{\rm SDW}$ 
or in the metallic phase as observed in magnetic Raman scattering and neutron 
scattering  \cite{Ishikado,Matan,Diallo2010,Sugai,Okazaki}.  
It strongly suggests that the superconductivity is induced by the same magnetic 
correlation as in the SDW state. 

The magnetism of iron pnictide cannot be simply understood by the localized 
or itinerant spin model, because the magnetic moment $M=0.8\sim 0.9\mu_B/{\rm Fe}$ 
is intermediate \cite{Goldman,Kaneko,Ewings,Matan,Huang,Diallo2009,Johannes}.  
The SDW phase is usually stabilized by the decrease of the electronic energy 
by opening of the gap at the Fermi energy in the folded bands.  
In the iron pnictide the SDW is not simply explained by the energy gain due to 
the gap formation induced by the Fermi surface nesting between the hole pocket 
at the $\Gamma$ point and the electron pocket at the $M$ point in $k$ space.  
The spin structure is consistent with the Fermi surface 
nesting in BaFe$_2$As$_2$ \cite{Mazin,Kuroki,Dong,Ding,Terashima}, but not in 
FeTe, because the spin stripe direction is different by 45$^{\circ}$ \cite{Li,Bao}.   
Johannes and Mazin \cite{Johannes} pointed out that the energy gain is mainly 
one-electron energy balance for the antiferromagnetic (AF) spin patterns.  
The energy gain by the nesting is only a part of it.  
In fact a full SDW gap does not open in BaFe$_2$As$_2$ as 
known from the fact that the electric conductivity in the SDW state is even 
better than in the normal state. 
It is caused by the remaining Dirac nodes \cite{Wang,Ran,Harrison,Morinari}.  
Furthermore the magnetic exchange interaction energies are not simply obtained 
from the conventional short-range superexchange mechanism \cite{Yildirim,Si,Yin,Han,Zhao}.  
The exchange interactions are very different, if they are obtained from 
the total energy of the long-range stripe spin structure.  
The exchange interaction in the $y$ direction is antiferromagnetic in the former 
case, but ferromagnetic in the latter case.

BaFe$_2$As$_2$ undergoes the SDW state by the first order 
transition \cite{Dong2,Huang,Chu2009,Kitagawa,Baek,Tanatar,Diallo2010} 
accompanied by the tetragonal $(I4/mmm)$-orthorhombic $(Fmmm)$ 
structural transition at $T_{\rm SDW}=137$ K \cite{Rotter,Cruz,Huang}.  
The magnetic order in the SDW state is a stripe type in which nearest-neighbor 
spins are antiparallel in the $x$ direction and parallel 
in the $y$ direction \cite{Cruz,Kaneko,Ewings,Zhao,McQueeney,Matan}.  
In this article we use the $a$ and $b$ axes as the tetragonal crystallographic 
axes and the $x$ and $y$ axes as the orthorhombic axes which are rotated by 
45$^{\circ}$ from the $a$ and $b$ axes.  

The electronic states in the SDW state is different from the conventional one.  
Usually the SDW state is insulating, because a full SDW gap opens at the Fermi 
energy.  
Despite the SDW state, BaFe$_2$As$_2$ is metallic because electronic 
states remain at the Fermi energy ($E_{\rm F}$).  
It is expressed by the selective inter-orbital coupling in the multi-orbital 
bands of BaFe$_2$As$_2$ \cite{Kuroki,Qi,Raghu,Graser}.  
Wang {\it et al}. \cite {Wang} and Ran {\it et al}. \cite {Ran} predicted 
the nodal gaps composed of point-contact Dirac corns.  
The theoretical works of the Dirac nodes were 
reported \cite {Wang,Ran,Harrison,Morinari,Eremin,Andersen} and some of them 
showed that the Dirac nodes are protected \cite {Wang,Ran,Harrison,Morinari}.  
Angle-resolved photoemission spectroscopy (ARPES) observed the Dirac nodes 
near $E_{\rm F}$ \cite{Richard}.  
The $k$-linear dispersion of the Dirac corn is expressed by the massless 
relativistic Weyl equation \cite{Morinari} which expresses the motion of a 
neutrino.  
It is known that the two-dimensional massless Dirac Fermion exhibits a variety 
of unusual phenomena \cite{Novoselov}.  
For example, the interaction to the acoustic phonon is very weak, causing a 
very large mobility.  

In order to find out whether the gap excitations near the Dirac node and the 
anti-node were separately detected, the orbital components in the electron and 
the hole bands were calculated in the two-dimensional two-orbital tight band model.  
The most area of the band in the $k_x$-$k_y$ space is composed of 
the mixed $xz$ and $yz$ orbitals, but the electronic states along four lines are 
composed of pure $xz$ or $yz$ orbital.  
The same orbital components in two bands couple to open the SDW gap.  
The optical transition at the anti-node is from pure $xz$ to pure $xz$ or from 
pure $yz$ to pure $yz$, while that at the Dirac node is from pure $xz$ to pure 
$yz$ or from pure $yz$ to pure $xz$.  
The symmetry of the electronic transition at the anti-node is 
$A_{\rm 1g}$+$B_{\rm 2g}$ and that at the node is $B_{\rm 1g}$.  
The symmetry difference enables us to detect the anti-node and the node separately.  
The experimentally obtained $B_{\rm 2g}$ spectra show critical fluctuation just 
above $T_{\rm SDW}$ and jump into the gap structure at $T_{\rm SDW}$, while the 
$B_{\rm 1g}$ spectra gradually change as temperature 
decreases through the $T_{\rm SDW}$ \cite{SugaiSDWgap,Chauviere2}.  
The gap structure is composed of the reduced intensity below 300 cm$^{-1}$ and 
the peaks at 400 and 800 cm$^{-1}$.  
The anti-nodal gap energies are the same as those in 
infrared spectroscopy \cite{Hu} and ARPES \cite{Yang,Zhang,Liu,Fink,YiSDW,Yi}.  

The magnetic exchange interaction energies also cannot be simply determined.  
The exchange interaction energies calculated in the superexchange interaction 
model are antiferromagnetic with the similar magnitude in the $x$ and $y$ 
directions, $J_{1x} \approx J_{1y}$, because the anisotropy between $x$ and $y$ is 
small even in the orthorhombic SDW structure \cite{Yildirim,Si}.  
In order to stabilize the stripe spin structure, the next nearest neighbor 
exchange interaction energy $J_2$ must be larger than half of the nearest neighbor 
exchange interaction energy.  
On the other hand the exchange interaction energies obtained from the second 
derivative of the total energy of the stripe spin structure with respect to the 
angle of the moments are antiferromagnetic in the $x$ direction and weak 
ferromagnetic in the $y$ direction. \cite{Yin, Han}.  
The superexchange interaction is related to the strong correlation of on-site 
Hubbard Coulomb repulsion.  
It was argued that the systems are moderately correlated and the largely local 
iron moments are driven by Hund's intra-atomic exchange rather than by the 
on-site Hubbard repulsion \cite{Johannes}.  

The magnetic excitations observed by neutron scattering are analyzed (1) by the 
antiferromagnetic interactions in both $x$ and $y$ directions 
\cite{Ewings,Zhao,McQueeney,Lester,Diallo2009,Diallo2010}
and (2) antiferromagnetic in $x$ and ferromagnetic in $y$ \cite{Ewings,Zhao2}.  
Usually the full spin wave dispersion is obtained in the localized spin picture.  
The similar dispersion is also obtained in the itinerant spin picture, but the 
high-energy parts are decayed into particle-hole excitations (Stoner continuum) 
and the continuum excitations extend above the maximum of the spin wave 
dispersion \cite{Kariyado,Brydon,Kaneshita,Knolle}.  
Many low-energy neutron scattering experiments did not reach the top of the 
dispersion \cite{Ewings,Zhao,McQueeney,Matan,Lester,Diallo2010}.  
Diallo {\it et al}. \cite{Diallo2009} observed broad magnetic dispersion 
above 100 meV and suggested the damping of spin waves by the 
particle-hole excitations, while Zhao {\it et al}. \cite{Zhao2} reported 
the full spin wave dispersion expressed by the antiferromagnetic in $x$ and 
ferromagnetic in $y$.  
No consensus is obtained whether the itinerant or localized view is 
appropriate at present.    

In order to clarify whether the exchange interaction energies are determined 
by the short-range superexchange interaction mechanism or the total energy 
of the long-range stripe spin structure, the selection 
rule of two-magnon Raman scattering from the stripe spin structure 
was investigated.  
Two-magnon scattering in the insulating antiferromagnet is caused by the change 
of two spins $\Delta S^z=\pm 1$.  
We found that the two-magnon scattering mechanism can be separately considered 
for the change of spins at the nearest neighbor sites and the next nearest 
neighbor sites.  
They give different symmetries and peak energies.  
Applying this selection rule to the two-magnon Raman scattering, it is found 
that the exchange interaction are given by the total energy mechanism.  

The peak energy of the present metallic magnetic Raman spectra is the same 
as the two-magnon scattering from localized spins in insulator, but the large 
spectral weight above twice the maximum energy of the spin wave dispersion 
cannot be understood by the two-magnon scattering process.  
Magnetic Raman spectra in metal is usually expressed by the 
two-magnon scattering from decayed spin waves.  
The decrease of the spin correlation length reduces the magnetic excitation 
energy as well as the life time.  
Therefore the large spectral weight at high energy cannot be interpreted by 
this model.  
We have to take into account the magnetic component of the itinerant carriers 
travelling in the antiferromagnetic spin sea.  
The effect is interpreted both in the localized spin model and the itinerant 
spin model.
In the localized spin model the high-energy magnetic scattering is interpreted as 
electronic scattering of itinerant carriers in the AF spin lattice, 
because the moving carriers overturn spins and the electron spectral function 
have the magnetic component expressed by the self-energy in the string 
model \cite{Brinkman,Shraiman,Marsiglio,Dagotto,Martinez,Liu2,Lee,Manousakis}.  
In the itinerant model the magnetic scattering is interpreted as particle-hole 
excitations from the majority spin states to the minority spin states 
\cite{Kariyado,Brydon,Kaneshita,Knolle}.  
The magnetic scattering in metallic BaFe$_2$As$_2$ is compared to the 
hole-doped La$_{2-x}$Sr$_x$CuO$_4$ and electron-doped Nd$_{2-x}$Ce$_x$CuO$_4$.  
The electron density dependence in BaFe$_2$As$_2$ is found to be different 
from the cuprates

The present experiments are based on the theoretical findings that the 
electronic excitations near the Dirac node and the anti-node can be separately 
detected using the different Raman symmetries and the nearest and the next 
nearest neighbor exchange interaction energies can be also separately detected 
using the different Raman symmetries.  
Hence the relevant theories are presented followed by the experimental results.  
Section 2.1 presents the electronic Raman scattering mechanism 
of the SDW gap which is different from the normal metal and the superconducting gap.  
Section 2.2 presents the formulation of the nodal gap and anti-nodal gap in the 
two-orbital tight binding model.  
Section 2.3 presents the experimental procedure.  
Section 2.4 presents the experimental results of the high-energy spectra of the 
electronic Raman scattering and the magnetic Raman scattering.  
Section 2.5 presents the low-energy spectra of the Dirac node and the anti-node.  
Section 2.6 presents phonons.  
Section 3.1 presents the robust antiferromagnetic correlation in two-dimensional 
magnetism.  
Section 3.2 presents the symmetry of two-magnon scattering.  
Section 3.3 presents the experimental results of two-magnon Raman scattering.  
Section 3.4 presents the effect of conductive carriers to the magnetic scattering in 
comparison to the cuprate superconductors.

\section{SDW State}
\subsection{\label{sec:Raman}Raman scattering of the SDW gap}
\begin{figure}
\begin{center}
\includegraphics[trim=0mm 0mm 0mm 0mm, width=8cm]{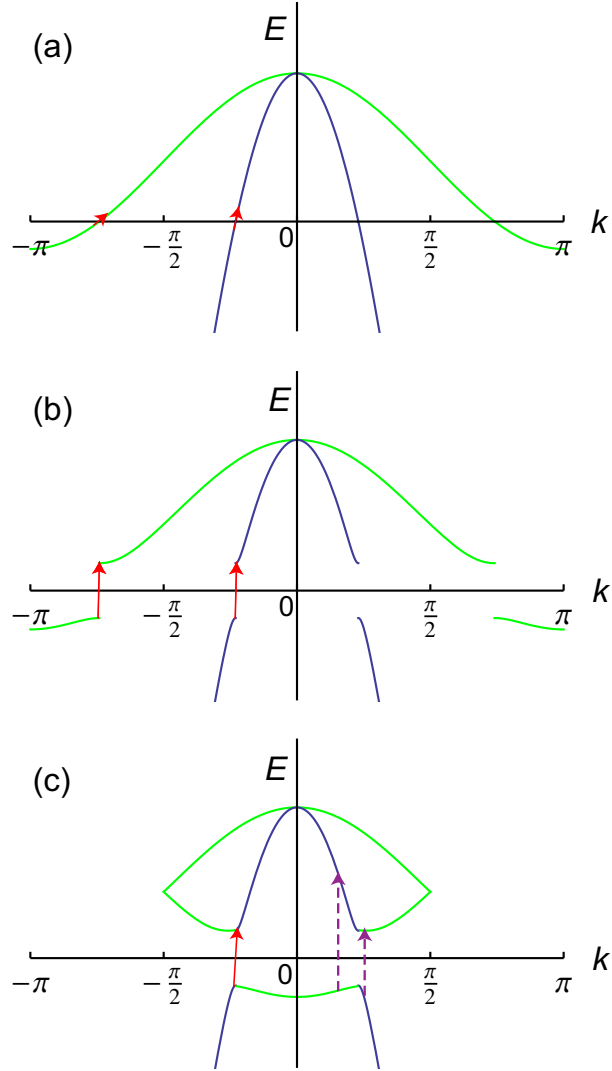}
\caption{(color online) 
(a) Intraband electronic Raman excitations (red arrows) in the normal state.  
The green band has the electron pocket at $k=\pi$ and the blue band has 
the hole pocket at $k=0$.  
The momentum shift in the Raman process is kept to be small in order to 
satisfy the momentum conservation with incident and scattered light.  
(b) In the superconducting state the excitation across the gap gives the 
pair breaking spectra.  
The electronic transition is limited to near the original Fermi surface.  
The initial and the final states are on the same band.  
(c) In the SDW state the electronic bands are folded into $-\pi/2\le k\le \pi/2$.  
The intraband electronic excitations are limited to near the band crossing parts, 
while the interband transitions (purple dashed arrows) are available anywhere.  
Therefore the interband transition is dominant.  
}
\label{fig1}
\end{center}
\end{figure}
Sometimes the SDW gap is treated in the similar way as the superconducting gap 
in the Raman scattering experiment.  
However, their excitation processes are completely different.  
The Raman scattering of the superconducting gap is expressed by the intraband 
transition with the momentum shift 
$|{\bm k}_{\rm f}-{\bm k}_{\rm i}|=\Delta k \approx 0$ at 
${\bm k}_{\rm i}\approx {\bm k}_{\rm F}$, where ${\bm k}_{\rm F}$ is the Fermi 
wave vector.  
On the other hand the SDW gap scattering is expressed by the interband transition 
between the original band and the folded band produced by the momentum shift of 
${\bm k}_{\rm nest}$, where ${\bm k}_{\rm nest}$ is the nesting vector $(\pi, 0)$.  
The vertical transition is allowed all over the folded Brillouin zone.  

Electronic Raman scattering in the normal state is induced by the first order 
of the $\bm A^2$ term and the second order of the $\bm p\cdot \bm A$ term in the 
electron-radiation interaction term $(\bm P-\frac{e}{c}\bm A)^2$.  
The matrix element is \cite{Wolff,Platzman}
\begin{eqnarray}
M=e^{\alpha}_{\rm i}e^{\beta}_{\rm s} \frac{1}{m} \left[\delta_{\alpha \beta} 
+\frac{1}{m}\left(\sum_{t} \frac{<c,{\bm k}_{\rm f}|P_{\beta}|b,{\bm k}_{\rm i}+{\bm q}_{\rm i}><b,
{\bm k}_{\rm i}+{\bm q}_{\rm i}|P_{\alpha}|a,{\bm k}_{\rm i}>}
{\epsilon_{t,{\bm k}+{\bm q}_{\rm i}}-\epsilon_{s,{\bm k}}-\omega_{\rm i}}+X\right) \right]
\label{eq:matrix},
\end{eqnarray}
where $X$ is the term with the different time order, $m$ the free electron mass, 
$e^{\alpha}_{\rm i}$ and $e^{\beta}_{\rm s}$ polarization vectors of incident and 
scattered light, ${\alpha}$ and $\beta$ the Cartesian coordinate, 
$\omega_{\rm i}$ and ${\bm q}_{\rm i}$ the incident photon energy 
and wave vector, $a$, $b$, and $c$ are the bands of the initial, intermediate, 
and final electronic states, and ${\bm k}_{\rm i}$ and ${\bm k}_{\rm f}$ are the initial and 
final wave vectors of the electron.  
The scattered photon wave vector ${\bm q}_{\rm s}$ is 
${\bm k}_{\rm i}+{\bm q}_{\rm i}-{\bm k}_{\rm f}$.  
The $q_{\rm i}$ and $q_{\rm s}$ are approximately zero.

If the electron returns to the initial band ``{\it intraband transition}, $c=a$", 
eq.~(\ref{eq:matrix}) is the same form as the ${\bm k}\cdot {\bm p}$ perturbation.  
In the low energy and the long wavelength approximation of the incident light 
eq.~(\ref{eq:matrix}) becomes 
\begin{eqnarray}
M=e^{\alpha}_{\rm i}e^{\beta}_{\rm s} \left(\frac{1}{m^*}\right)_{\alpha \beta}
\label{eq:mass},
\end{eqnarray}
where $1/m^*$ is the effective inverse mass tensor at ${\bm k}_{\rm i}$.  

If the final band $c$ is different from the initial band $a$ ``{\it interband 
transition}, $c\ne a$", the scattering probability is determined by the 
symmetries of the initial and final electron wave functions.  

Figure 1 shows the electronic Raman excitations in (a) the normal metal, 
(b) superconducting state, and (c) SDW state.  
In the normal state the momentum shift across the Fermi wave vector is the 
order of $\Delta k\lesssim k_{\rm ZB}/1000$ to satisfy the momentum conservation 
with incident and scattered light, where $k_{\rm ZB}$ is the wave vector at the 
Brillouin zone boundary.  
The energy shift in Raman scattering from the intraband transition (red arrows) 
is usually limited to less than a few tens cm$^{-1}$, if the strong correlation 
effect is not taken into account.  
The scattering probability is expressed using the inverse mass tensor of 
eq.~(\ref{eq:mass}).  

The electronic transition in the superconducting state is the 
{\it intraband transition}.  
The electronic excitation is limited to near the gap because of the momentum 
conservation with light.  
The scattering probability is expressed using the inverse mass tensor of 
eq.~(\ref{eq:mass}) \cite{Klein,Cardona,Sugai,MazinGap}.  
The superconducting gap symmetry with the largest scattering intensity in 
BaFe$_{1.84}$Co$_{0.16}$As$_2$ is $B_{\rm 2g}$ \cite{Muschler,Sugai,Chauviere2}.  
The scattering intensities from the electron pockets and hole 
pockets were calculated using the inverse mass tensors and the screening terms in 
two-band model \cite{Sugai} and the full-potential linearized augmented plane wave 
(LAPW) model \cite{MazinGap}.  
The importance of the multi-orbital was pointed \cite{Moreo,Sugai,Zhang2}

In the SDW state the electronic bands are folded into $-\pi/2\le k\le \pi/2$ for 
the nesting vector ${\bm k}_{\rm nest}=(\pi, 0)$.  
The {\it intraband transition} is limited to near the gap as 
in the superconducting gap.  
However, the dominant electronic excitations are the {\it interband transition} 
(purple dashed arrows) 
between the original band and the folded band all over the folded Brillouin zone.  
Among them the transitions at the rather wide area near the gap make the gap 
structure.  
The Raman intensity of the SDW gap is one hundred times larger than that of the 
superconducting gap.  
The electronic Raman scattering of the {\it interband transition} is not given by 
the inverse mass tensor.  
The symmetries of the wave functions directly result in the scattering 
probability.  
Then we have to know the nested band structure to understand the 
Raman scattering of the SDW gap.

\subsection{\label{sec:two-band}SDW state in the two-band model}
The electronic states near the Fermi energy is created from the two-dimensional 
FeAs layer.  
Wang {\it et al}. \cite{Wang} and Ran {\it et al}. \cite{Ran} pointed out 
that the SDW gap in BaFe$_2$As$_2$ is 
not a full gap, but a nodal gap.  
In the two-orbital band model the hole pocket changes the dominant orbitals 
$xz$-$yz$-$xz$-$yz$ on circling the Fermi surface, while the electron 
Fermi surface is represented by only one orbital.  
Under the conditions of (1) the collinear SDW order (denoted by the 
magnetization direction of $x$), (2) the inversion symmetry, and (3) the 
time-reversal symmetry, the vorticity is $\pm 2$ for the hole pockets and 
0 for the electron pockets.  
It creates the Dirac nodes with the vorticity $\pm 1$ in the SDW gap.  
The nodes are located at the intersection between the SDW wave vector and 
the hole Fermi surface in the two-band model.  
The Dirac nodes are shown to be stable in the five band model, too \cite{Ran,Morinari}.  

The different optical selection rule near the node and the 
anti-node is expressed by the two-orbital model of $xz$ and $yz$.  
Using the transfer integrals by Qi {\it et al.} \cite{Qi} and 
Raghu {\it et al}. \cite{Raghu} 
the tight-binding Hamiltonian is 
\begin{eqnarray}
H_{0}=\sum_{\bm{k},\sigma}(d^\dagger_{xz,\sigma}(\bm{k}),d^\dagger_{yz,\sigma}
(\bm{k})){\bm K}(\bm{k})
\left(\begin{array}{c}
d_{xz,\sigma}(\bm{k}) \\d_{yz,\sigma}(\bm{k})
\end{array}\right)\;,
\end{eqnarray}
where $2 \times 2$ matrix ${\bm K}({\bm k})$ is 
\begin{eqnarray}
&&{\bm K}({\bm k})=(\epsilon_+({\bm k})-\mu)\tau_0+\epsilon_{xy}({\bm k})\tau_1+
\epsilon_-({\bm k})\tau_3,\nonumber\\
&&\epsilon_+({\bm k})=-(t_1+t_2)(\cos k_x +\cos k_y)-4t_3\cos k_x\cos k_y,\nonumber\\
&&\epsilon_-({\bm k})=-(t_1-t_2)(\cos k_x-\cos k_y),\nonumber\\ 
&&\epsilon_{xy}({\bm k})=-4t_4\sin k_x \sin k_y,
\end{eqnarray}
where $d^\dagger_{xz,\sigma}$ and $d_{xz,\sigma}$ are the creation and annihilation 
operators of a $d$ electron with the 
$xz$ orbital and spin $\sigma$, $\mu$ the chemical potential, $\tau_i$ are 
Pauli matrices, and $\tau_0$ is the unit matrix.  
$t_1$ is the transfer integral between $xz$ orbitals and $t_2$ between $yz$ orbitals 
at the nearest neighbor sites in the $x$ direction.  
$t_3$ is the transfer integral between $xz$ orbitals and $t_4$ between $xz$ and $yz$ 
orbitals at the diagonal sites.  
The band dispersions are given by
\begin{eqnarray}
E_{\pm}({\bm k})=\epsilon_+({\bm k})\pm \sqrt{\epsilon^2_-({\bm k})+\epsilon^2_{xy}({\bm k})}-\mu.
\end{eqnarray}
The upper and lower bands are called the electron band and the hole band, 
because they make the electron pockets and the hole pockets, respectively.  
The eigen functions are
\begin{eqnarray}
\Psi_{\pm,\sigma}({\bm k})&=&\frac{\epsilon_{xy}({\bm k})}{A_{\mp}}\ d_{xz,\sigma}\nonumber\\
& &+\frac{-\epsilon_-({\bm k})\pm \sqrt{\epsilon^2_-({\bm k})+
\epsilon^2_{xy}({\bm k})}}{A_{\mp}}\ d_{yz,\sigma},\nonumber\\
\label{eq:eigen}
\end{eqnarray}
where the normalization factors $A_{\pm}$ are
\begin{eqnarray}
A_{\pm}=\sqrt{2\left[\epsilon^2_{xy}({\bm k})+\epsilon^2_-({\bm k})\pm \epsilon_-({\bm k}) \sqrt{\epsilon^2_-({\bm k})+\epsilon^2_{xy}({\bm k})}~ \right]}.\nonumber\\
\end{eqnarray}
The $xz$ and $yz$ components on the electron and hole bands are given 
by eq.~(\ref{eq:eigen}).  

\begin{figure}
\begin{center}
\includegraphics[trim=0mm 0mm 0mm 0mm, width=15cm]{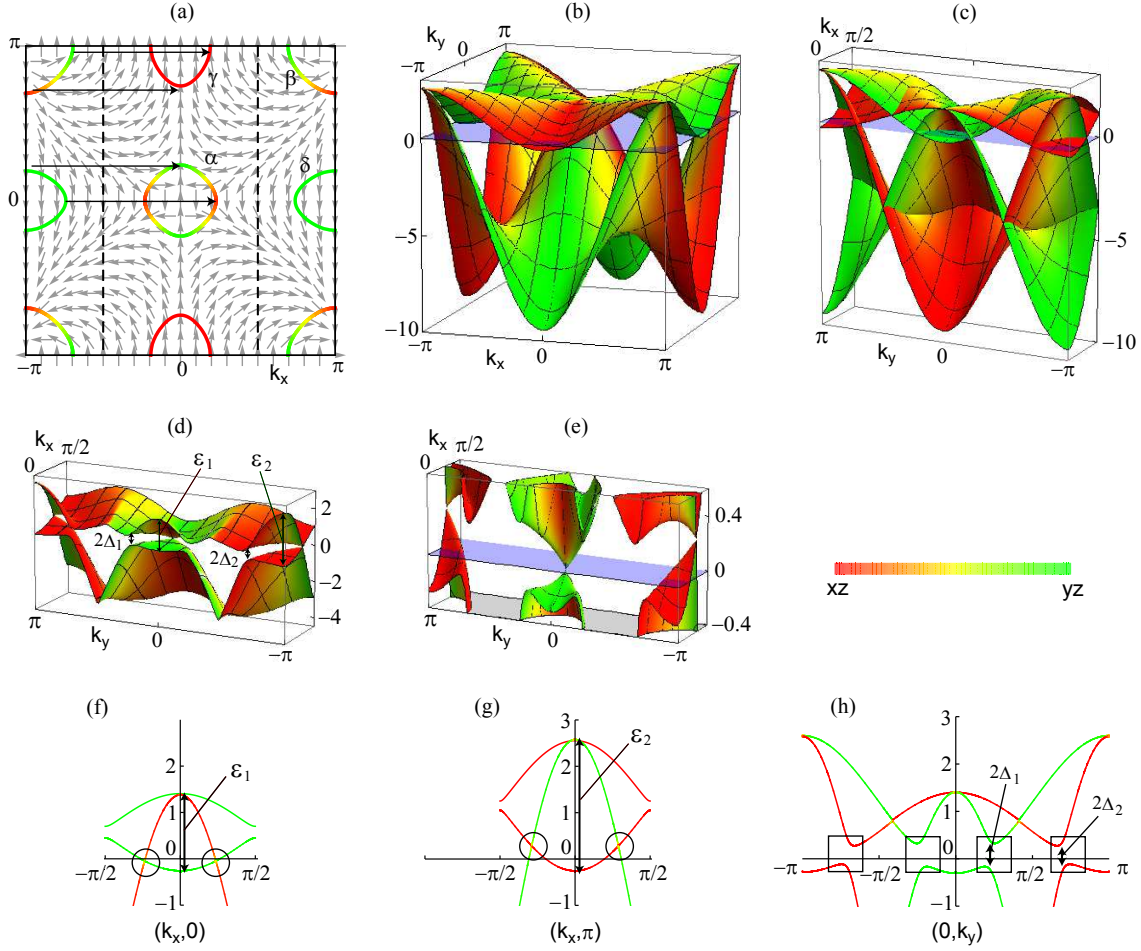}
\caption{(color online) 
(a) Hole Fermi surfaces $\alpha$ and $\beta$ and electron Fermi surfaces 
$\gamma$ and $\delta$.  
The $xz$ and $yz$ orbital components are shown by red and green according to 
the color scale.  
The gray arrows indicate the unit vectors parallel to $(-\epsilon_{xy}(\bm{k}), 
\epsilon_{-}(\bm{k}))$ \cite{Qi,Raghu}.  
The dashed lines shows the unit cell in the SDW state.  
(b) Dispersions of the electron and hole bands.  
The $E_{\rm F}$ is shown by the translucent blue plane.  
(c) Folded bands in the SDW state.  
(d) Nodal gapped bands near the $E_{\rm F}$.  
The uppermost dispersion is removed.  
(e) The same as above in the expanded energy scale.  
Note that the front side is $E-k_x$ in (b), while $E-k_y$ in (c)-(e).  
(f)-(h) Dispersions in $(k_x, 0)$, $(k_x, \pi)$, and $(0, k_y)$.  
The circle denotes the Dirac node and the square denotes the anti-node.  
}
\label{fig2}
\end{center}
\end{figure}
Figure 2(a) shows the 
Fermi surfaces of hole pockets $\alpha$ and $\beta$ and electron pockets $\gamma$ and 
$\delta$ in the Brillouin zone of one Fe atom per unit cell.  
The transfer integrals are $t_1=-1$, $t_2=1.3$, $t_3=t_4=-0.85$ in the unit of 
$|t_1|$ \cite{Raghu}.  
The gray arrows are unit vectors in the direction of 
$(-\epsilon_{xy}(\bm{k}),~ \epsilon_{-}(\bm{k}))$, 
where $\epsilon_{xy}(\bm{k})$ represents the mixing 
between $xz$ and $yz$ orbitals and $\epsilon_-(\bm{k})$ 
represents the difference of $xz$ and $yz$ component \cite{Qi,Raghu,Graser,Graser2010}.  
The upward component of the gray arrow is the $xz$ component (red) and the 
downward component is the $yz$ component (green) for the electron band and 
opposite for the hole band.  
The hole Fermi surface is composed of two orbitals whose weights gradually 
change on going round the Fermi surface.  
The Fermi surfaces in the $k_x$ and $k_y$ directions from (0, 0) are composed of the 
pure $xz$ and $yz$ orbitals, respectively.  
The SDW state is formed by shifting the Brillouin zone by ${\bm k}_{\rm nest}=(\pi, 0)$.  
The reduced zone into $-\pi/2\le k_x \le \pi/2$ is shown by the dashed lines.  
Figure 2(b) shows the dispersions of the electron band and the hole band.
The unit of the energy is $|t_1|$.  
The weight of the $xz$ and $yz$ orbitals are represented by the scale from 
red to green.  
The translucent blue plane shows the $E_{\rm F}$.  
Figure 2(c) shows the dispersion in a half of the folded Brillouin zone 
without the band mixing.  
The crossing parts may split by the mixing.  
Note that the front side of the cross section is the $E-k_y$.  

The local interaction to form the SDW with the wave vector $(\pi, 0)$ is \cite{Ran}
\begin{eqnarray}
H_{\rm SDW}&=&M_{ab}\sum_i(-1)^{i_x}(d^\dagger _{i,a\uparrow}d_{i,b\uparrow}
-d^\dagger_{i,a\downarrow}d_{i,b\downarrow}), 
\end{eqnarray}
where 
\begin{eqnarray}
M_{ab}&=&[\phi_0\tau_0+\phi_1\tau_1+\phi_2\tau_2+\phi_3\tau_3]_{st}, 
\label{eq:int}
\end{eqnarray}
and $a$ and $b$ are orbitals and $i_x$ is the $x$ coordinate of the 
iron atomic site $i$ in the unit of the inter-iron atomic distance.  
The $(-1)^{i_x}$ term gives the twofold periodicity in the $x$ direction.  
$\phi_i$ is the mean field parameter to minimize the free energy 
including the correlation term.  
If the intra-orbital interaction $\tau_0$ or $\tau_3$ is dominant, the 
Dirac node appears at $k_y=0$ and $\pi$ which are the same as the calculations 
by Ran {\it et al.} \cite{Ran}, Morinari {\it et al.} \cite{Morinari}, and 
Kaneshita {\it et al.} \cite{Kaneshita,KaneshitaPRL}.

Figures 2(d) and 2(e) show the gapped two bands near the $E_{\rm F}$.  
The same orbital components in two bands couple to open a gap, if the $\tau_0$ or 
$\tau_3$ is dominant.  
The difference is scarcely found whether $\tau_0$ or $\tau_3$ is dominant.  
The connecting point of two Dirac corns is a little below $E_{\rm F}$ near (0, 0) 
and above $E_{\rm F}$ near $(0, \pi)$, which causes small electron Fermi surfaces 
near $(0, 0)$ and hole Fermi surfaces near $(0, \pi)$.  
The electronic transitions near the Dirac nodes appear at the energy near zero 
and those at the anti-node appear at $2\Delta_1$ and $2\Delta_2$.  
The gap energies are $2\Delta_1=0.49$ and $2\Delta_2=0.39$, when the 
coupling constant $\phi_0=0.3$ or $\phi_3=0.3$ in the energy unit of $|t_1|$.  
The transitions near the Dirac node and anti-node have different symmetries, 
because the former is the transition between $yz$ and $xz$ while the 
latter between $yz$ and $yz$ near (0, 0) or $xz$ and $xz$ near $(\pi, 0)$.  
The $\Delta k\approx 0$ transition $\epsilon_1$ from $yz$ to $xz+yz$ at (0, 0) and 
$\epsilon_2$ from $xz$ to $xz+yz$ at $(0, \pi)$ in Fig. 2(d) have large joint 
density of states.  
The energies are not strongly affected by the coupling term eq.~(\ref{eq:int}).  
The dispersions in $(k_x, 0)$, $(k_x, \pi)$, and $(0, k_y)$ are shown in 
Figs. 2(f)-(h).

\subsection{Experimental Procedure}
Single crystals of BaFe$_2$As$_2$ were grown by the self-flux method.  
Raman spectra were measured on the fresh cleaved surfaces in a quasi-back 
scattering configuration using 5145 \AA \ laser light.  
The polarization configuration of the Raman spectra is presented by 
$(e_{\rm i} e_{\rm s})$.  
The crystallographic axes of the tetragonal structure are $a$ and $b$.  
The bisecting directions are $x$ and $y$.  
The $x$ and $y$ directions are rotated by 45$^{\circ}$ from the $a$ and $b$.  
The observed Raman spectra include phononic, magnetic, and electronic excitations.  
Raman active phonon modes are $1A_{\rm 1g}+1B_{\rm 1g}+2E_{\rm g}$ in the 
tetragonal structure \cite{Litvinchuk,Rahlenbeck,Choi,Chauviere}.  
The orthorhombic structure ($Fmmm$) is the sub-group of the tetragonal 
structure ($I4/mmm$).  
The second order phase transition is allowed from the change of the symmetry, 
but whether the transition is the first order or the second order is 
determined by the free energy.  
It is known that the transition becomes the first order, if the SDW 
transition is accompanied by the structural phase transition.  
In the present case the first order transition of BaFe$_2$As$_2$ has been 
confirmed in many experiments \cite{Dong2,Huang,Chu2009,Kitagawa,Baek,Tanatar,Diallo2010}.  
The $(aa)$ spectra allow the $A_{\rm 1g}+B_{\rm 1g}$ ($A_{\rm g}+B_{\rm 1g}$) 
modes, $(ab)$ $B_{\rm 2g}$ ($A_{\rm g}$) mode, 
$(xx)$ $A_{\rm 1g}+B_{\rm 2g}$ ($A_{\rm g}$), and $(xy)$ $B_{\rm 1g}$ 
($B_{\rm 1g}$) in the tetragonal (orthorhombic) structure.  
The Raman system was calibrated using a black body radiation so that the 
intensity is proportional to the Raman probability independently to the 
scattered light wave number.

\subsection{\label{sec:Anti-nodal}High-energy spectra of the anti-nodal gaps 
and the transitions at $(0, 0)$ and $(0, \pi)$ points}
\begin{figure}
\begin{center}
\includegraphics[trim=0mm 0mm 0mm 0mm, width=7cm]{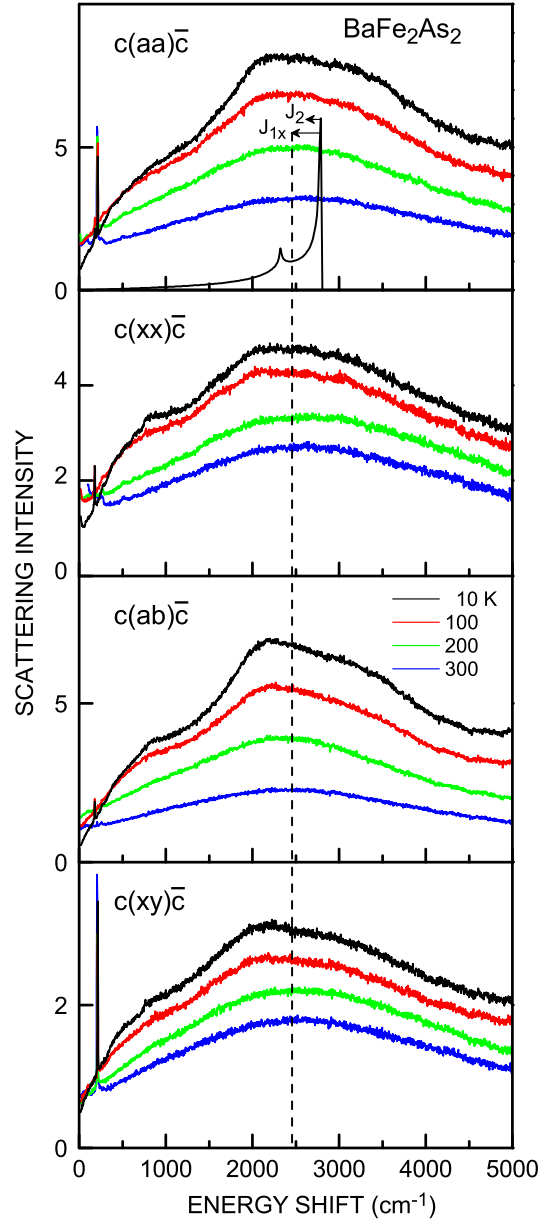}
\caption{(color online) 
Temperature dependence of polarized Raman spectra in BaFe$_2$As$_2$.  
}
\label{fig3}
\end{center}
\end{figure}
\begin{figure}
\begin{center}
\includegraphics[trim=0mm 0mm 0mm 0mm, width=7cm]{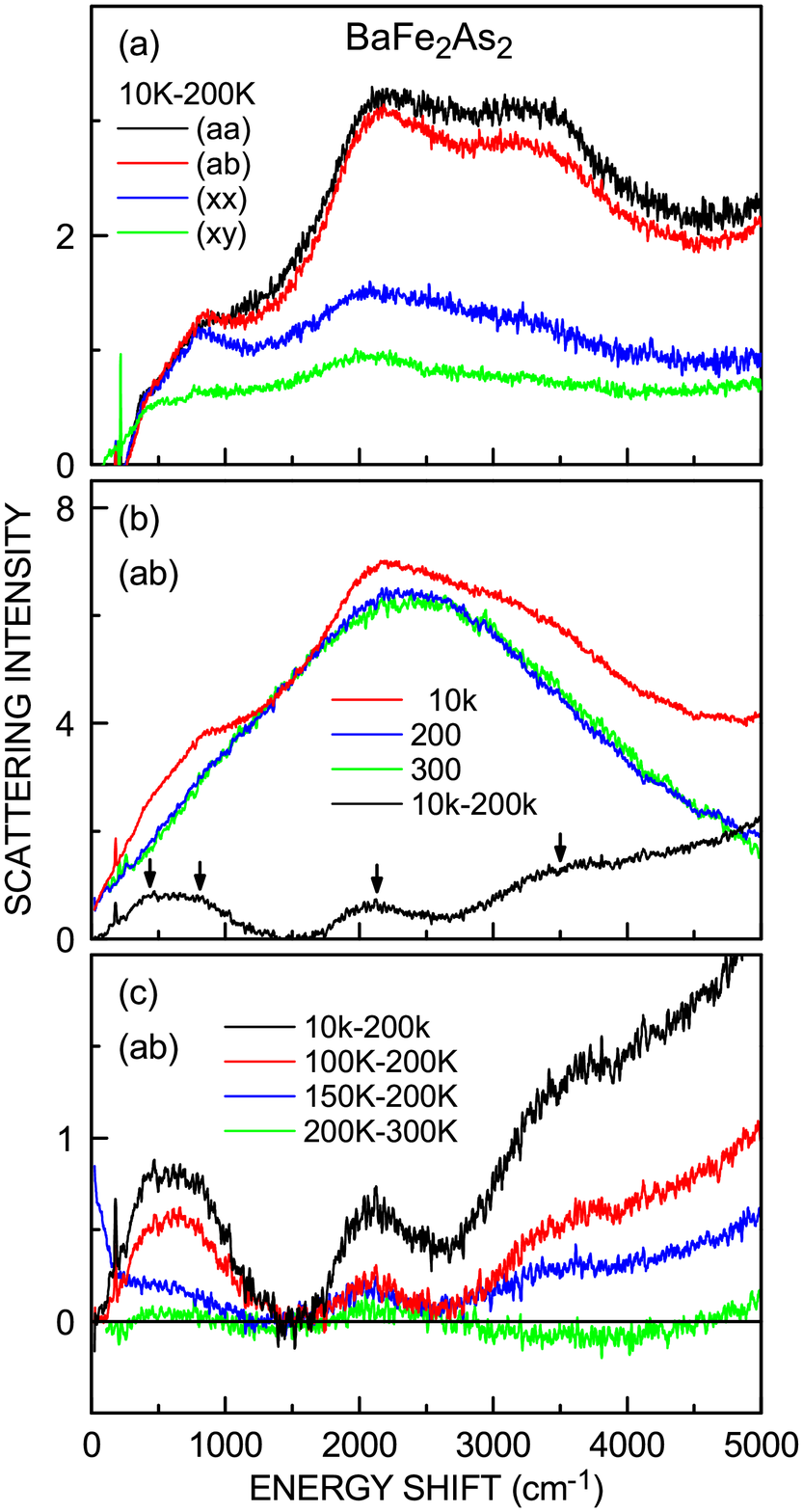}
\caption{(color online) 
(a) Differential Raman spectra between the SDW state at 10 K and the normal 
state at 200 K.  
(b) The Raman spectra at 10 K in the SDW phase and at 200 and 300 K in the normal 
phase.  
The 10 K spectra include electronic scattering activated by the Brillouin zone 
folding and magnetic scattering.  
The 200 and 300 K spectra is only magnetic scattering above a few tens cm$^{-1}$.  
The spectra at 200 K and 300 K are expanded in intensity and the zero levels are 
adjusted.  
The differential spectra between 10 K and 200 K are shown by the black curve.  
(c) Differential spectra obtained in the same way as (b).  
}
\label{fig4}
\end{center}
\end{figure}
Figure 3 shows the temperature dependence of polarized Raman spectra.  
The sharp peaks at 181 and 215 cm$^{-1}$ are caused by the $A_{\rm 1g}$ and 
$B_{\rm 1g}$ phonons, respectively.  
The broad hump with the top energy $2400\sim 2500$ cm$^{-1}$ at 300 K is the 
two-magnon scattering peak.  
The two-magnon intensity increases with decreasing temperature.  
Below $T_{\rm SDW}\approx 130$ K new peaks appear at 400, 800, 2150, and 3500 
cm$^{-1}$.  

As stated in \S~\ref{sec:Raman} the electronic scattering in the normal phase 
is limited to less than a few tens cm$^{-1}$.  
The wide-energy intensity is derived from only magnetic excitations.  
In order to estimate the interband transition activated by the SDW transition, 
the differential spectra between the SDW 
state at 10 K and the normal state at 200 K are shown in Fig. 4(a).  
The intensity scales are the same for different polarization configurations.  
It is noted that the differential spectra include the electronic scattering 
component which is activated in the SDW state and the two-magnon scattering 
component whose intensity increases with decreasing temperature.  
In order to clearly show the magnetic scattering component, the scattering 
intensities in the normal states at 200 K and 300 K are enlarged to the 
estimated level at 10 K in Fig. 4 (b).  
The zero levels are adjusted.  
The spectra at 200 K and 300 K are almost the same.  
It confirms that the spectral shape of the magnetic scattering is the same 
except the intensity in the normal phase.  
The extra humps appeared at 10 K are created by the interband electronic 
transitions.  
The differential spectra between 10 K and 200 K are 
shown by the black curve.  
The peak positions are shown by the arrows.  
The 800 and 400 cm$^{-1}$ peaks are assigned to the anti-nodal SDW gaps of 
$2\Delta_1$ and $2\Delta_2$ and the 2150 and 3500 cm$^{-1}$ peaks to the 
transitions $\epsilon_1$ at (0, 0) and $\epsilon_2$ at $(0, \pi)$ points as 
shown in Figs. 2(d), 2(f), and 2(g).  
These peak intensities are large in $(aa)$, $(ab)$, and $(xx)$ and small in 
$(xy)$.  
The $2\Delta_1$ and $2\Delta_2$ are the transition from $yz$ to $yz$ and from 
$xz$ to $xz$, respectively.  
The $\epsilon_1$ is the transition from $yz$ to $yz$ and $xz$.  
The density of the final state is larger in $yz$ than $xz$.  
The $\epsilon_2$ is the transition from $xz$ to $xz$ and $yz$.  
The density of the final state is larger in $xz$ than $yz$.  
The symmetry of the transitions from $xz$ to $xz$ and from $yz$ to $yz$
is $A_{\rm 1g}+B_{\rm 2g}$ and that from $xz$ to $yz$ is $B_{\rm 1g}$.  
The $(aa)$, $(xx)$, $(ab)$, and $(xy)$ spectra are active in 
$A_{\rm 1g}+B_{\rm 1g}$, $A_{\rm 1g}+B_{\rm 2g}$, $B_{\rm 2g}$, and $B_{\rm 1g}$, 
respectively.  
Therefore the $(xx)$, $(ab)$, and $(aa)$ spectra observe above four transitions 
strongly and $(xy)$ spectra observe $\epsilon_1$ and $\epsilon_2$ weakly.  
The experimental results roughly satisfy the selection rule.  

Figure 4(c) shows the differential spectra between the SDW phase and 
the normal phase at 200 K, and also between 200 K and 300 K in the normal phase.  
The 400, 800, 2150, and 3500 cm$^{-1}$ peaks start to increase below the 
$T_{\rm SDW}$.  
The peaks are observed at 150 K which is 
a little above $T_{\rm SDW}\approx 130$ K of this sample.  
The intensities of these peaks increase like the second order transition as 
temperature decreases.  

\begin{figure}
\begin{center}
\includegraphics[trim=0mm 0mm 0mm 0mm, width=7.5cm]{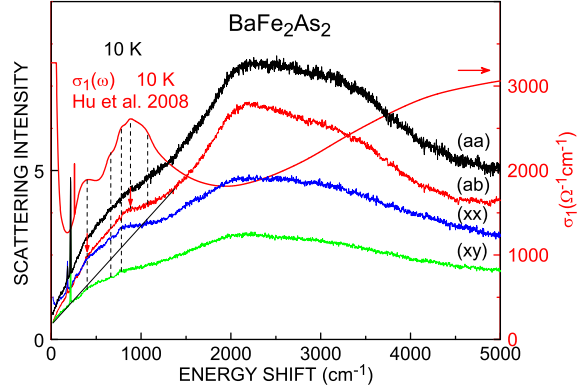}
\caption{(color online) 
Comparison between the Raman spectra and the optical conductivity spectra \cite{Hu} 
in the SDW state.  
}
\label{fig5}
\end{center}
\end{figure}
Figure 5 shows the 10 K spectra at various polarization configurations and 
the optical conductivity at 10 K obtained by Hu {\it et al}. \cite{Hu}. 
The 400 and 800 cm$^{-1}$ SDW gap peaks in the present Raman spectra correspond 
to the 360, 890 cm$^{-1}$ peaks in the optical conductivity \cite{Hu}.  
They are also correspond to the gaps observed in ARPES \cite{Richard}.  
The fine peaks are noticeable in Raman and infrared spectra.  
The 662 and 780 cm$^{-1}$ peaks in the $(xx)$ and $(xy)$ spectra and 
the 882 and 1072 cm$^{-1}$ peaks in the $(aa)$ and $(ab)$ spectra correspond to 
the fine structure in the optical conductivity spectra \cite{Hu}.  
There are two possibilities for the origin of the fine structure.  
One is the electronic transition and the other is the phonon side band. 
Many other transitions may appear in the five band model \cite{Ran,KaneshitaPRL}.  
Yi {\it et al}. \cite{YiSDW} reported in ARPES that the 
electronic structures are reconstructed in the SDW state and 
may not be described in the simple folding scenario by the correlation 
effect, because the introduction of Hubbard $U$ in the local-density 
approximation to fit the band dispersion changes the structure at $E_{\rm F}$.  
The continuum between 400 and 800 cm$^{-1}$ may have such effects.  
The Fermi surfaces have rather large three-dimensional characters \cite{Graser2010} and 
it may induce some complicated structure.  
On the other hand the energy difference 118 cm$^{-1}$ and 190 cm$^{-1}$ 
between the main peak 
and the fine peak is close to the 117 cm$^{-1}$ $E_{\rm g}$ phonon energy 
and the 182 cm$^{-1}$ $A_{\rm 1g}$ phonon energy \cite{Rahlenbeck}.  
Those phonon modes forming the phonon side bands may be localized 
modes \cite{Mahan}. 

The humps at 2150 and 3500 cm$^{-1}$ are not observed in the infrared spectra, 
because the $d$-$d$ transition is infrared inactive.  
We assigned the humps to the $\epsilon_1$ and $\epsilon_2$ transitions in Fig. 2(d).  
Those transitions may be related to the flat bands whose energies decrease with 
the increase of $U$.  
The $U$ reduces the magnetic moment $M=0.5\sim 1~ \mu_{\rm B}/{\rm Fe}$ in the 
local-density approximation plus Hubbard U (LDA+U) band calculation \cite{YiSDW}.  
However, it is difficult to compare the Raman peaks to the band calculation, 
because the bandwidth is renormalized and the energy shift is different at the 
$\Gamma$ and X points to fit the ARPES data.

\subsection{Low-energy spectra of the Dirac node and the anti-node}
\begin{figure*}
\begin{center}
\includegraphics[trim=0mm 0mm 0mm 0mm, width=15cm]{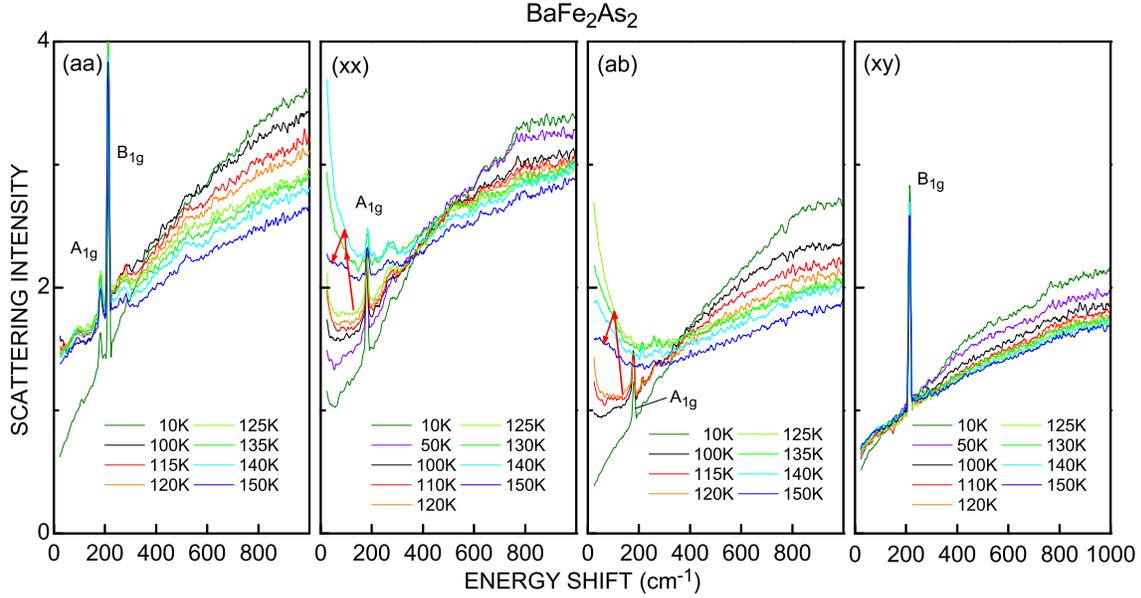}
\caption{(color online) 
Temperature dependent low-energy Raman spectra in BaFe$_2$As$_2$.  
The low-energy spectra below 300 cm$^{-1}$ jump to the high-intensity level 
at $T_{\rm SDW}$ and then gradually decreases as temperature increases in the 
$(xx)$ and $(ab)$ spectra.  
The red arrows show the jumps.  
The $(aa)$ and $(xy)$ spectra do not show the jump.  
}
\label{fig6}
\end{center}  
\end{figure*}
\begin{figure*}
\begin{center}
\includegraphics[trim=0mm 0mm 0mm 0mm 0mm, width=15cm]{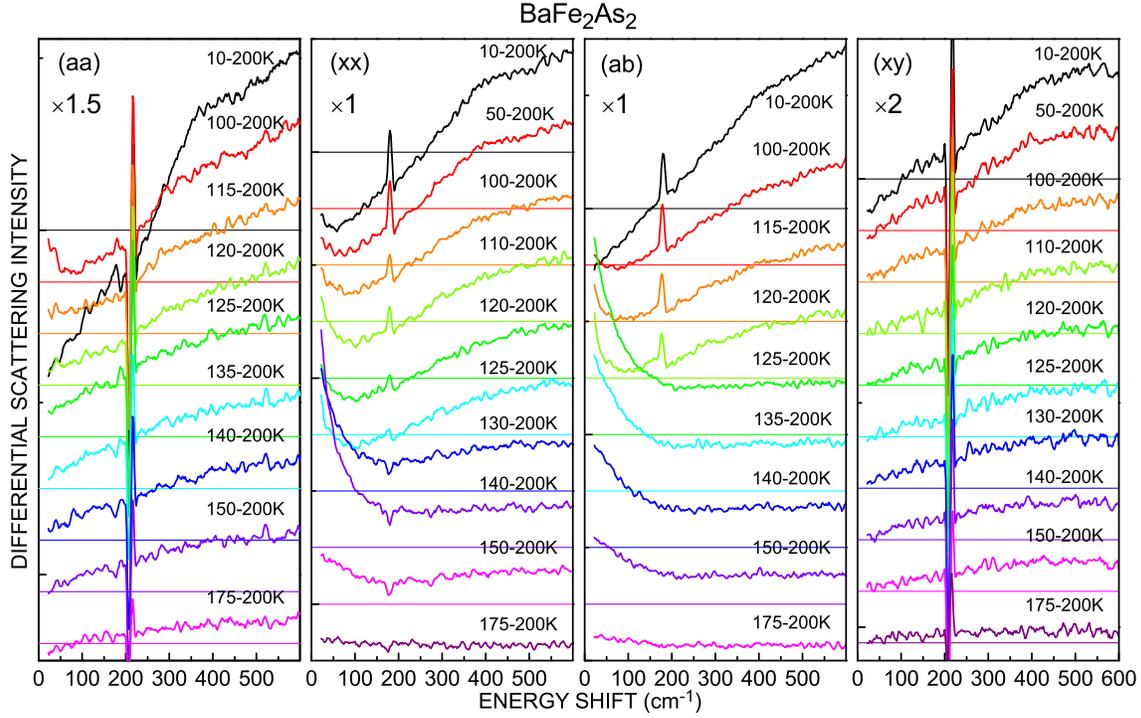}
\caption{(color online) 
Differential spectra from 200 K.  
The intensity below 200 cm$^{-1}$ increases in the $(xx)$ and $(ab)$ spectra 
as temperature decreases below 150 K and jumps to the low intensity at $T_{\rm SDW}$.  
The SDW-induced anti-nodal gap peaks at 400 and 800 cm$^{-1}$ gradually 
increase in intensity as temperature decreases below a little above $T_{\rm SDW}$ 
in all polarization configurations, but is weakest in $(xy)$.  
The zero levels are shown by the same color lines.
}
\label{fig7}
\end{center}  
\end{figure*}
The temperature dependence of the polarized low-energy Raman spectra 
is shown in Fig. 6.  
All the spectra are plotted in the same intensity scale.  
The sharp peaks at 181 and 215 cm$^{-1}$ are the $A_{\rm 1g}$ and $B_{\rm 1g}$ 
phonons in the tetragonal phase, respectively.  
They are discussed later.  
The 400 and 800 cm$^{-1}$ humps are observed at 10 K in all polarization 
configurations.  
However, the temperature dependence below 300 cm$^{-1}$ is very different 
between the $(aa)$ and $(xy)$ group and the $(xx)$ and $(ab)$ group.  
In the $(xx)$ and $(ab)$ spectra the low-energy scattering intensity abruptly 
increases from 125 K to 130 K in the $(xx)$ spectra and from 120 K to 125 K 
in the $(ab)$ spectra, and then gradually decreases as temperature 
increases \cite{SugaiSDWgap,Chauviere2}.  
On the other hand the jump is not observed in the $(aa)$ and $(xy)$ spectra 
near the $T_{\rm SDW}$.  
The small temperature difference of the jump in $(xx)$ and $(ab)$ is induced 
by the first order transition.  

It is noted that the experiment was made in the sequence of $(aa)$ and 
$(ab)$ at 10 K, $(aa)$ and $(ab)$ at 100K, $\cdots$, 300 K and after rotating 
the sample by 45$^{\circ}$ $(xx)$ and $(xy)$ at 10 K, $(xx)$ and $(xy)$ at 50 K, 
$\cdots$, 300 K.
The $(aa)$ and $(ab)$ spectra at the same temperature are obtained exactly 
in the same condition except for the scattered light polarization.  
The same is for $(xx)$ and $(xy)$.  
Therefore the spectral jump at $T_{\rm SDW}$ in the $(xx)$ and $(ab)$ spectra 
and the continuous change in the $(aa)$ and $(xy)$ spectra are intrinsic 
properties of this material.  

Figure 7 shows the differential Raman spectra between various temperatures and 
200 K.  
The zero levels are shown by the horizontal lines with the same color as the 
spectra.  
In the $(xx)$ and $(ab)$ spectra the intensity below about 200 cm$^{-1}$ 
increases with decreasing temperature from 150 K to $T_{\rm SDW}\approx 130$ K 
and then jumps to decrease by the opening of the SDW gap.  
On the other hand the jump is not observed in the $(aa)$ and $(xy)$ spectra.  

The above difference between the $(xx)$ and $(ab)$ group and the $(aa)$ and 
$(xy)$ group is caused by the different symmetries for the electronic 
transitions near the anti-node and the Dirac node.  
The electronic transition near the anti-node is from $xz$ to $xz$ or from $yz$ 
to $yz$ and that near the Dirac node is from $xz$ to $yz$ or from $yz$ to $xz$.  
The symmetry of the transition near the anti-node is $A_{\rm 1g}+B_{\rm 2g}$ 
and that near the node is $B_{\rm 1g}$.  
The $(xx)$ spectra are active to $A_{\rm 1g}+B_{\rm 2g}$ and $(ab)$ to $B_{\rm 2g}$, 
$(aa)$ to $A_{\rm 1g}+B_{\rm 1g}$, and the $(xy)$ to $B_{\rm 1g}$.  
Therefore the $(xx)$ and $(ab)$ spectra represent the electronic transitions 
near the anti-node and the $(aa)$ and $(xy)$ spectra represent the Dirac node, 
if the $A_{\rm 1g}$ component from the anti-node is ignored.  
The spectra representing the anti-node have the jump into the gap spectra.  
On the other hand the spectra representing the Dirac node has not the specific 
change at $T_{\rm SDW}$.  
The $A_{\rm 1g}$ component might be observed in the $(aa)$ spectra, but the 
spectra do not show the jump at $T_{\rm SDW}$.  
Instead the $(aa)$ spectra gradually change into the spectra of the gapped 
excitations.  
In contrast to the $(aa)$ spectra the gap spectra are weak in the $(xy)$ spectra 
which represents only the excitation near the node.  

The increase of the low-energy scattering intensity on approaching the 
$T_{\rm SDW}$ from high temperature is induced by the critical fluctuation 
relating to the opening of the anti-nodal gap, because it is observed only in 
the $B_{\rm 2g}$ symmetry in which the anti-nodal gap excitations are observed.  
The magnetic fluctuation above $T_{\rm SDW}$ was also observed in neutron 
scattering \cite{Matan}.  
The critical fluctuation is the characteristic properties of the second order 
phase transition.  
The simultaneous SDW and structural transitions in BaFe$_2$As$_2$ splits into two 
second-order-like transitions on substituting Co for Fe \cite{Chu2009}. 
The phase locking of the SDW to the lattice potential by the simultaneous 
structural transition changes the transition into the first order and the SDW 
into the commensurate one.  
The SDW and structural phase transition of BaFe$_2$As$_2$ is, however, close to 
the second order transition.  
It induces the critical fluctuation.  

The above analysis is done on the assumption that the dominant orbital mixing 
term is $\tau_0$ or $\tau_3$ in eq.~(\ref{eq:int}).  
If $\tau_1$ or $\tau_2$ is dominant, the positions of the Dirac node and the 
anti-node are exchanged and the observed spectral symmetries at the $T_{\rm SDW}$ 
become inconsistent to those of the electronic transitions.  

As discussed above the symmetry dependence of the observed spectra is consistently 
interpreted by the electronic excitations near the anti-node and the Dirac node.  
However, the dominant orbitals at the nesting parts in electron pocket may be 
modified in the five band model \cite{Kuroki,Graser,Cao,Graser2010,Ikeda,Andersen}.
More detailed theoretical analysis is expected.

\subsection{Phonon mode activated in the SDW state}
The sharp peaks in the tetragonal phase above $T_{\rm SDW}$ are observed at 
181 ($A_{\rm 1g}$) cm$^{-1}$ in the $(aa)$ and $(xx)$ spectra and 215 cm$^{-1}$ 
($B_{\rm 1g}$) in the $(aa)$ and $(xy)$ spectra of
Fig. 6 \cite{Litvinchuk,Rahlenbeck,Choi,Chauviere}.  
Below $T_{\rm SDW}$ the ($A_{\rm 1g}$) peak appears at 187 cm$^{-1}$ in the 
$(ab)$ spectra.  
The $A_{\rm 1g}$ mode in the tetragonal ($I4/mmm$) structure changes into the 
$A_{\rm g}$ mode in the orthorhombic structure by the compatibility relation 
of the group theory.  
Taking into account the 45$^{\circ}$ rotation of the crystallographic axes at 
the phase transition from the tetragonal to the orthorhombic structure, the Raman 
intensity of the $A_{\rm g}$ mode in the $(ab)$ (in the tetragonal notation) 
polarization configuration is proportional to $|R_{11}-R_{22}|^2$ 
using the Raman tensor in the orthorhombic structure.  
The intensities in the $(aa)$, $(xx)$, and $(yy)$ spectra are proportional to 
$\frac{1}{4} |R_{11}+R_{22}|^2$, $|R_{11}|^2$, and $|R_{22}|^2$, respectively.  
The $(xx)$ and $(yy)$ spectra may mix if the crystal is twinned, but both 
spectra represents the same $A_{\rm g}$ mode and the difference is small even 
in the untwinned crystal.  
Other polarization configurations are not affected by the twin structure.  
The intensities in the $(aa)$, $(xx)$, and $(yy)$ spectra are expected to be 
nearly the same and much larger than that in the $(ab)$ spectra because 
$R_{11}\approx R_{22}$.  
Instead at 10 K the intensities in the $(xx)$ and $(ab)$ spectra are 
$1.5\sim 2$ times as large as the intensity in $(aa)$.   
This phonon is the mode in which As atom moves in the $c$ 
direction \cite{Litvinchuk}.  
It has large magneto-phonon interaction, because the Fe-As-Fe angle is very 
sensitive to the Fe-Fe exchange interaction energy \cite{Yildirim,Yin,Kuroki2}.  
The detailed mechanism of the enhancement is still an open question.  
The enhancement of the $A_{\rm 1g}$ phonon below $T_{\rm SDW}$ was reported 
in CaFe$_2$As$_2$ \cite{Choi} and BaFe$_{2-x}$Co$_x$As$_2$ \cite{Chauviere}.  
Similar enhancement of the infrared active phonon was reported in 
BaFe$_2$As$_2$ \cite{Akrap}

\section{Magnetic Raman Scattering}

\subsection{Antiferromagnetic correlation in low-dimensional magnetism}
The exchange interaction energies are very different whether the calculation 
starts from the short-range superexchange interaction or from the total energy 
of long-range AF stripe spin structure.  
In the former case the calculation of the local superexchange interaction 
gives AF exchange interaction energies for all directions 
reflecting the equivalent $x$ and $y$ directions.  
The stripe spin order is stable if $J_{1x}=J_{1y}<2J_2$, otherwise 
the checkerboard spin order is stable \cite{Yildirim,Si}.  
On the other hand in the latter case the exchange interaction energies are 
obtained from $J_{ij}(R)=-\partial ^2E/\partial \theta _i(0) \partial \theta _j(R)$, 
where $E$ is the total energy and $\theta _j(R)$ is the angle of the moment 
of the $j$th spin \cite{Yin, Han}.  
The calculated energies are antiferromagnetic $SJ_{1x}=43$, ferromagnetic 
$SJ_{1y}=-3.1$, and antiferromagnetic diagonal $SJ_2=14.3$ meV \cite{Han}.  
The low-energy spin wave in the SDW state was observed by neutron scattering 
in BaFe$_2$As$_2$ \cite{Ewings,Matan}, SrFe$_2$As$_2$ \cite{Zhao} 
and CaFe$_2$As$_2$ \cite{McQueeney}.  
Ewings {\it et al}. \cite{Ewings} fitted the spin wave velocity at 7 K in 
BaFe$_2$As$_2$ by (1) the long-range exchange interaction 
model with $SJ_{1x}=36$, $SJ_{1y}=-7.2$, and $SJ_2=18$ meV and (2) the short-range 
exchange interaction model with $SJ_{1x}=17,5$, $SJ_{1y}=17.5$, and $SJ_2=35$ meV.  
Zhao {\it et al}. disclosed in high-energy neutron scattering that the entire 
spin wave dispersion of CaFe$_2$As$_2$ ($T_{\rm N}\approx 170$ K) at 10 K is 
expressed by the long-range model with $SJ_{1x}=49.9$, $SJ_{1y}=-5.7$, $SJ_2=18.9$, 
and $SJ_z=5.3$ meV \cite{Zhao2}.  

The spin wave dispersion ignoring the anisotropy terms is given by \cite{Ewings}
\begin{equation}
\hbar \omega(k)=\sqrt{A^2-D^2}
\label{eq:spinwave},
\end{equation}
\begin{equation}
A=2S \left\{ J_{1y}\left[ \cos \frac{k_y}{2}-1\right]+J_{1x}+2J_2+J_z\right\},
\end{equation}
\begin{equation}
D=2S\left\{ J_{1x}\cos \frac{k_x}{2}+2J_2\cos \frac{k_x}{2} \cos \frac{k_y}{2}+J_z\cos k_z \right\}.
\end{equation}
\begin{figure}
\begin{center}
\includegraphics[trim=0mm 0mm 0mm 0mm, width=7cm]{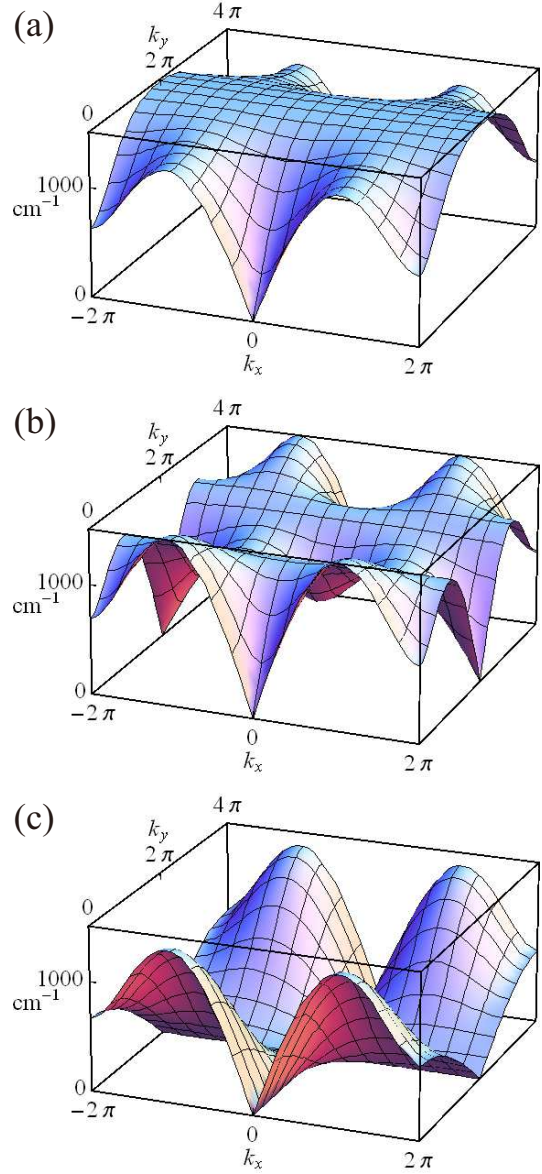}
\caption{(color online) 
Spin wave dispersion at $k_z=0$ with (a) $SJ_{1x}=36$, $SJ_{1y}=-7.2$, 
$SJ_2=18$ \cite{Ewings}, and $SJ_z=5.3$ meV, (b) $SJ_{1x}=17.5$, $SJ_{1y}=17.5$, 
$SJ_2=35$ \cite{Ewings}, and $SJ_z=5.3$ meV, and (c) $SJ_{1x}=40$, $SJ_{1y}=40$, 
$SJ_2=20$, and $SJ_z=5.3$ meV.  
}
\label{fig8}
\end{center}
\end{figure}
Figure 8 shows the dispersion at $k_z=0$ with (a) $SJ_{1x}=36$, $SJ_{1y}=-7.2$, 
$SJ_2=18$ \cite{Ewings}, and $SJ_z=5.3$ meV, (b) $SJ_{1x}=17.5$, $SJ_{1y}=17.5$, 
$SJ_2=35$ \cite{Ewings}, and $SJ_z=5.3$ meV, and (c) $SJ_{1x}=40$, $SJ_{1y}=40$, 
$SJ_2=20$, and $SJ_z=5.3$ meV.  
(a) and (b) have nearly the same spin wave velocity, but 
the maximum energy in (b) is lower than in (a).  
The energy at $k_x=0$ or $k_y=0$ goes to zero at the critical point of (c) 
$J_{1x}=J_{1y}=2J_2$ in Fig. 8(c).  

The superconducting phase does not overlap the SDW phase or slightly overlaps 
at the crossover region, if it is observed by quasi-static experiments such as 
magnetic susceptibility, resistivity, specific heat, $\mu$SR, and crystal 
structure \cite{Chu2009,Ni,Drew,Fernandes}.  
Sometimes it is called as the evidence of competing superconductivity 
and magnetism.  
However, such a phase diagram is valid only at $k\approx 0$ and $\omega \approx 0$.  
Neutron scattering experiments in high temperature superconductors disclosed 
that only the magnetic correlation perpendicular to the CuO$_2$ plane 
disappears above the magnetic transition temperature \cite{Ishikado,Matan,Diallo2010}.  
This is a common property in low dimensional magnetism.  
The AF transition temperature $T_{\rm N}$ in two-dimensional antiferromagnet 
is determined by the weakest exchange interaction $J_z$ as \cite{Thio}
\begin{equation}
T_{\rm N}=J_z\left(\frac{L}{d}\right)^2, 
\end{equation}
where $L$ is the intralayer AF spin correlation length and $d$ is the inter-spin 
length in the layer.  
Neutron scattering showed that the correlation length is $15\sim 18$ \AA\ at 
136 K ($\sim T_{\rm SDW}$) and it remains to be $6\sim 8$ \AA\ at 180 K 
\cite{Matan,Diallo2010}.  
The magnetic excitations with the wavelength longer than the correlation length 
may disappear but those with shorter wavelength survive above $T_{\rm N}$ and even 
in the superconducting phase.  
Two-magnon Raman scattering in the AF insulator represents the joint density of 
states of $k$ and $-k$ magnons to satisfy the momentum conservation with light.  
The dominant part is derived from the zone boundary with the large density of states.  
Therefore the two-magnon peak survives till the last as the correlation length 
decreases.

\subsection{\label{sec:2mag}Symmetry of two-magnon scattering induced 
by Coulomb interaction in insulator}
Two-magnon scattering in AF insulator arises from the simultaneous changes of 
$\Delta S_z=\pm 1$ on both spin sublattices by the Coulomb 
interaction \cite{Moriya,Fleury,Fleury1968,Elliott,Elliott2,Parkinson}.  
The two-magnon Hamiltonian can be written as \cite{Fleury}
\begin{eqnarray}
H_{\rm two-mag}=\sum_{\begin{subarray}{c}\bm{Rr}\\{\alpha \beta \gamma \delta} 
\end{subarray}}B_{\alpha \beta \gamma \delta}(\bm{r})e_{\rm i}^{\alpha}e_{\rm s}^{\beta}S_{\bm{R}}^{\gamma}S_{\bm{R}+\bm{r}}^{\delta},
\end{eqnarray}
where $e_{\rm i}$ and $e_{\rm s}$ are incident and scattered light 
polarizations and $\bm{r}$ is the vector from the site $\bm{R}$ to the 
antiparallel spin site.  
$B$ is determined so that $H_{\rm two-mag}$ is totally symmetric.  
The spin term $S_{\bm{R}}^{\gamma}S_{\bf{R}+\bm{r}}^{\delta}$ is taken so that 
the total $z$ component of excited spins is zero as 
\begin{eqnarray}
S_{\bm{R}}^{x}S_{\bm{R}+\bm{r}}^{x}+S_{\bm{R}}^{y}S_{\bm{R}+\bm{r}}^{y}=\frac{1}{2} (S_{\bm{R}}^+S_{\bm{R+r}}^-+S_{\bm{R}}^-S_{\bm{R+r}}^+)
\label{eq:three}.
\end{eqnarray}

\begin{figure}
\begin{center}
\includegraphics[trim=0mm 0mm 0mm 0mm, width=7cm]{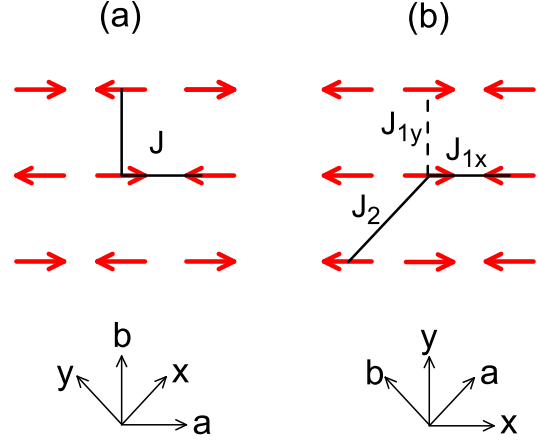}
\caption{(color online) 
Spin structures in (a) the checkerboard type in the cuprate superconductors and 
(b) the stripe type in BaFe$_2$As$_2$.  
The AF exchange interactions are presented by the line segments and 
the ferromagnetic interaction by the dashed line segment.  
Note that $a$ and $b$ are the tetragonal crystallographic axes at high temperatures.  
}
\label{fig9}
\end{center}
\end{figure}
In order to clarify the difference in the selection rule of the 
stripe AF spin structure in BaFe$_2$As$_2$ and the checkerboard AF spin structure 
in cuprate superconductors, the checkerboard-type is first discussed.  
The spin structure is shown in Fig. 9(a).  
The nearest neighbor spins are in the $a$ and $b$ directions.  
All the nearest neighbor spins are antiparallel.  
The next nearest neighbor spins are parallel and cannot contribute to 
two-magnon scattering.  
The next nearest neighbor exchange interaction is much smaller than the 
nearest neighbor exchange interaction.  
The system Hamiltonian is 
\begin{eqnarray}
H=J\sum_{kl}\bm{S}_k\cdot \bm{S}_l.
\end{eqnarray}
The two-magnon scattering Hamiltonian is given by \cite{Elliott,Elliott2,Parkinson}
\begin{eqnarray}
H_{\rm two-mag}=\sum_{kl}A(\bm{e}_{\rm i}\cdot \bm{r}_{kl})(\bm{e}_{\rm s}\cdot \bm{r}_{kl})(\bm{S}_k\cdot \bm{S}_l),
\label{eq:2mcb}
\end{eqnarray}
where $\bm{r}_{kl}$ is the unit vector connecting sites $k$ and $l$ sites.  
Here the spin longitudinal term $S_{\bm{R}}^{z}S_{\bm{R}+\bm{r}}^{z}$ 
is added to eq.~(\ref{eq:three}).  
Two-magnon scattering is active in $(aa)$ and $(xy)$ and inactive in $(ab)$.  
In $(xx)$ the two-magnon scattering Hamiltonian is the same as the system 
Hamiltonian except for the proportionality constant and the two-magnon 
scattering is inactive, because the two-magnon Hamiltonian commutes to the 
system Hamiltonian.  
The two-magnon peak energy is lower than the peak energy in the density of 
independent two magnon states by the magnon-magnon interaction energy $J$, 
because the exchange interaction energy between the nearest neighbor overturned 
spins is the same as before.

Figure 9(b) shows the stripe spin structure in BaFe$_2$As$_2$.  
It should be noted that the nearest neighbor directions are $x$ and $y$ 
in BaFe$_2$As$_2$ instead of $a$ and $b$ in cuprate.
The $a$ and $b$ axes are the tetragonal crystallographic axes.  
The nearest neighbor spin arrangement has two-fold rotational symmetry, 
antiparallel in the $x$ direction and parallel in the $y$ 
direction.  
All the second nearest neighbor spins are antiparallel.  
The system Hamiltonian is 
\begin{eqnarray}
H=&&J_{1x}\sum_{\bm{R},\bm{r}=\pm \bm{x}}\bm{S}_{\bm{R}}\cdot \bm{S}_{\bm{R}+\bm{r}}
+J_{1y}\sum_{\bm{R},\bm{r}=\pm \bm{y}}\bm{S}_{\bm{R}}\cdot \bm{S}_{\bm{R}+\bm{r}}\nonumber\\
&&+J_2\sum_{\bm{R},\bm{r}=\pm \bm{x}\pm \bm{y}}\bm{S}_{\bm{R}}\cdot \bm{S}_{\bm{R}+\bm{r}}
\end{eqnarray}
The nearest and the nest nearest neighbor spins should be taken into account, 
because both exchange interaction energies are the same order, for example, 
$SJ_{1x}=36$ meV and $SJ_2=18$ meV \cite{Ewings}.  
The two-magnon scattering Hamiltonian from the nearest neighbor spins is given by 
\begin{eqnarray}
H_{\rm two-mag}^{\rm nn}=\sum_{\bm{r}=\pm \bm{x}}C\bm{e}_{\rm i}^x\bm{e}_{\rm s}^x
(S_{\bm{R}}^+S_{\bm{R}+\bm{r}}^-+S_{\bm{R}}^-S_{\bm{R}+\bm{r}}^+), 
\end{eqnarray}
because only antiferromagnetic spins in the $x$ direction can contribute to 
the scattering.  
Two-magnon scattering is active, if both incident and scattered light has the 
electric field component in the $x$ direction.  
Then allowed polarizations are $(xx)$, $(aa)$, and $(ab)$.  
The crystals have magnetic twin structure for the interchange of $x$ and $y$ axes.  
The $(yy)$ and $(xx)$ polarizations cannot be distinguished.  
Then two-magnon scattering is inactive only in $(xy)$.  
The magnon-magnon interaction energy for the nearest neighbor spin exchange is 
$J_{1x}$.  

The two-magnon scattering Hamiltonian from the second nearest neighbor spins 
are the same as eq.~(\ref{eq:2mcb}) of the checkerboard spin structure.
The two-magnon scattering is active in $(xy)$ and $(aa)$, and inactive in $(ab)$.  
The $(xx)$ polarization is active, 
because the two-magnon Hamiltonian no longer commutes to the system Hamiltonian.  
The magnon-magnon interaction energy for the second neighbor spin exchange is $J_2$.  

The sum of the nearest and next nearest neighbor terms gives two-magnon intensity.  
Therefore it is active in all polarization configurations.  
In the case of antiferromagnetic $J_{1x}$ and ferromagnetic $J_{1y}$ the two-magnon 
scattering energy without the magnon-magnon interaction is the same in all polarization 
configurations, because the spin wave energy keeps constant along $k_y=2\pi$ in Fig. 8(a).  
The magnon-magnon interaction reduces the peak energy from twice the maximum energy 
of the density of states by $J_{1x}$ in $(ab)$ and by $J_2$ in $(xy)$, 
because only the nearest neighbor gives scattering in $(ab)$ and the next 
nearest neighbor gives scattering in $(xy)$.  
In the $(aa)$ and $(xx)$ spectra both components contribute.

\subsection{Two-magnon scattering in BaFe$_2$As$_2$}
\begin{figure}
\begin{center}
\includegraphics[trim=0mm 0mm 0mm 0mm, width=7.5cm]{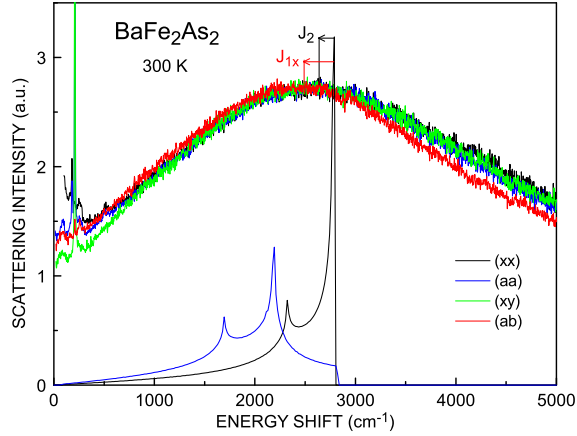}
\caption{(color online) 
Polarization configuration dependence of the two-magnon peak.  
The sharp curves are the density of independent two magnon states calculated 
using eq.~(\ref{eq:spinwave}) with the parameters $J_{1x}=36$, $J_{1y}=-7.2$, 
$J_2=18$, and $J_z=0$ meV (black) and $J_{1x}=17.5$, $J_{1y}=17.5$, $J_2=35$, 
and $J_z=0$ meV (blue).  
Two-magnon scattering peak energy is lower than the peak energy in the 
density of independent two magnon states by the 
exchange interaction energy between overturned spins.  
The $(ab)$ and $(xy)$ peak energies are smaller by $J_{1x}$ and $J_2$, 
respectively.  
The $(aa)$ and $(xx)$ are intermediate between them.  
}
\label{fig10}
\end{center}
\end{figure}
Figure 10 shows the polarized Raman spectra in the normal state at 300 K.  
The broad peak at 2500 cm$^{-1}$ is the two-magnon peak.  
In order to compare the two-magnon peak energies at 300 K in different polarization 
configurations, the intensities are adjusted so that the peak heights are 
the same.  
The spin wave densities of states calculated using eq.~(\ref{eq:spinwave}) 
with $J_{1x}=36$, $J_{1y}=-7.2$, $J_2=18$, and $J_z=0$ meV (black) and $J_{1x}=17.5$, 
$J_{1y}=17.5$, $J_2=35$, and $J_z=0$ meV (blue) are shown in Fig. 10.  
The energy is doubled from that of the density of single spin wave states.  
It is the density of independent two magnon states.  
The peak energy should be higher than the two-magnon scattering 
peak energy, because the magnon-magnon interaction reduces the two-magnon 
scattering peak energy.  
Therefore the blue density of states which is obtained from the superexchange 
interaction model is ruled out.  
The peak energy in the calculated density of independent two magnon states 
(black) is 2790cm$^{-1}$.  

The experimentally obtained two-magnon peak energy in the $(ab)$ spectra is 
lower than the independent two magnon peak energy (black) by $J_{1x}$ and that 
in the $(xy)$ spectra is lower by $J_2$.  
The reduced energy $J_{1x}$ and $J_2$ are the expected magnon-magnon 
interaction energies for the $(ab)$ and $(xy)$ spectra assuming $S=1$ as 
discussed at the end of \S~\ref{sec:2mag}.  
In the $(aa)$ and $(xx)$ spectra the decreased energy is the average 
of $J_{1x}$ and $J_2$.  
The experimentally obtained peak energy of two-magnon scattering is perfectly 
expressed by the total energy model of the long-range stripe spin structure.  
The very broad peak may be attributed to the short correlation length in 
the metallic state.  
However, the large weight at the higher energy than twice the highest 
spin wave energy may not be explained by the above Coulomb interaction-induced model.  
It is discussed in the next subsection.  

Very recently Chen {\it at al}. calculated two-magnon scattering peak energies \cite{Chentwomag}.  
The $B_{\rm 1g}$ peak energy is a little lower than the $A_{\rm 1g}+B_{\rm 2g}$ 
peak energies in the model with $J_{1y}=-0.1 J_{1x}$ and $J_2=J_{1x}$ in agreement 
with our result.  
They also showed that the $B_{\rm 1g}$ peak energy with $J_2=J_{1x}=J_{1y}$ is very low.  
The $B_{\rm 1g}$ spectra observe the spin wave in $k$ space with the weight of 
$(\cos k_x-\cos k_y)^2$.  
The energy decreases as $J_2$ approaches $J_{1x}/2=J_{1y}/2$ in Fig. 8(c).  
However, the $B_{\rm 1g}$ peak energy may not be so small, because the neutron 
scattering experiment shows $J_2\approx 2J_{1x}=2J_{1y}$ in the case 
of $J_{1x}=J_{1y}$ \cite{Ewings}.

\subsection{Magnetic excitations induced by moving carriers in the localized model}
\begin{figure*}
\begin{center}
\includegraphics[trim=0mm 0mm 0mm 0mm, width=15cm]{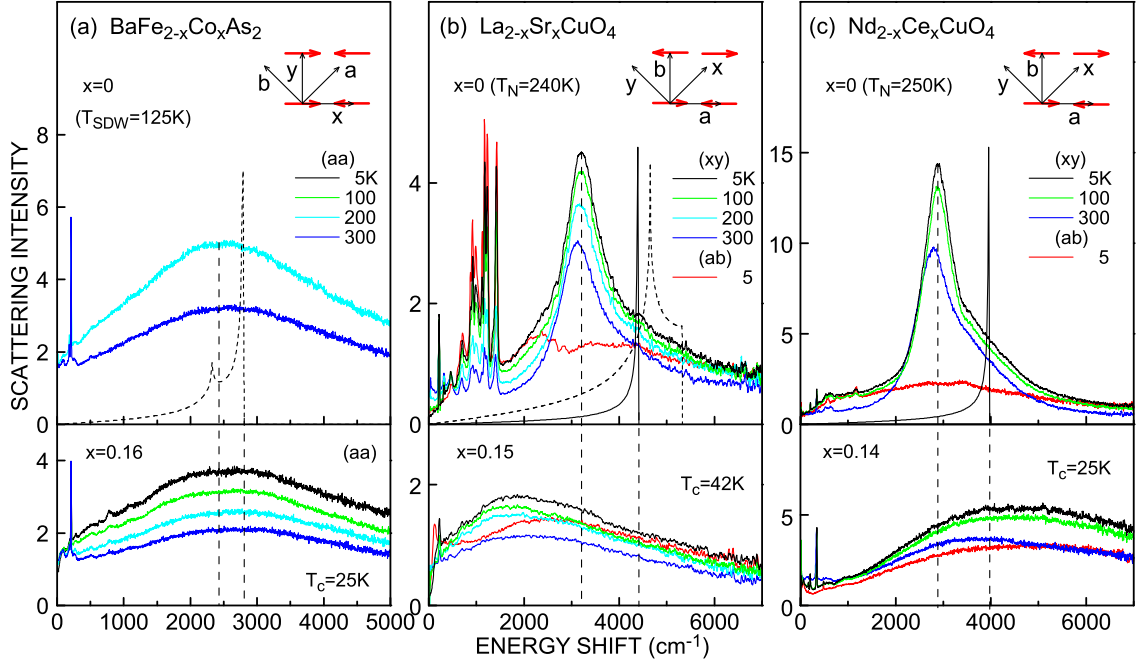}
\caption{(color online) 
(a) Temperature dependent polarized Raman spectra in the paramagnetic phase 
of BaFe$_2$As$_2$ (upper panel) and the superconductor 
BaFe$_{1.84}$Co$_{0.16}$As$_2$ (lower panel).  
The density of independent two magnon states is shown by the dashed line.  
The two-magnon peak energy and the independent two magnon excitation energy 
are shown by the dashed lines.  
The inset is the spin structure.  
(b) Two-magnon Raman spectra in the insulating La$_2$CuO$_4$ (upper panel) 
and the optimally hole-doped $(x=0.15)$ La$_{2-x}$Sr$_x$CuO$_4$ (lower panel).  
The density of independent two magnon states using single $J^*=118$ meV which is 
obtained from the two-magnon Raman scattering is shown by the solid line and 
that using $J^*=104$ and $J^*_2=-18$ meV which are obtained from neutron 
scattering \cite{Coldea} is shown by the dashed line in the upper panel.
The low-energy peak at $x=0.15$ in the $(ab)$ spectra of La$_{2-x}$Sr$_x$CuO$_4$ 
is the superconducting pair breaking peak.  
(c) Magnetic Raman spectra in the insulating Nd$_2$CuO$_4$ (upper panel) and 
the electron-doped superconducting Nd$_{2-x}$Ce$_x$CuO$_4$ (lower panel).  
The density of independent two magnon states with $J^*=855$ cm$^{-1}$ is shown 
by the solid line.  
}
\label{fig11}
\end{center}
\end{figure*}
The magnetic Raman scattering in cuprate superconductors are considered by 
the localized model.  
Figure 11 shows the comparison among (a) iron pnictides 
BaFe$_{2-x}$Co$_{x}$As$_2$, (b) hole-doped superconductor 
La$_{2-x}$Sr$_x$CuO$_4$, and (c) electron-doped superconductor 
Nd$_{2-x}$Ce$_x$CuO$_4$.  
Figure 11(a) shows the antiferromagnetic metal BaFe$_2$As$_2$ with 
$T_{\rm SDW}=130$ K and the superconductor 
BaFe$_{1.84}$Co$_{0.16}$As$_2$ with $T_{\rm c}=25$ K.  
The sharp dashed peak in the upper panel is the calculated density of independent 
two magnon states with $SJ_{1x}=36$, $SJ_{1y}=-7.2$, and $SJ_2=18$ meV. 
The two-magnon peak remains even in electron-doped superconductor 
BaFe$_{1.84}$Co$_{0.16}$As$_2$, indicating the short-range AF magnetic 
correlation remains \cite{Sugai}.  
The two-magnon peak is very broad and has large spectral weight above twice 
the maximum energy of the spin wave (right dashed line).  
The high-energy weight cannot be interpreted by the decayed magnon in the 
Coulomb interaction-induced two-magnon scattering process, because the 
higher-order multi-magnon scattering intensity is known to be weak \cite{Singh}.  

In La$_2$CuO$_4$ of Fig. 11(b) the sharp peaks from 700 to 1500 cm$^{-1}$ are 
two-phonon scattering peaks.  
The two-magnon scattering induced by the Coulomb interaction process 
is active in the $B_{\rm 1g}$ symmetry which is observed in the $(xy)$ spectra.  
The rather sharp peak around 3206 cm$^{-1}$ at 5 K in the insulating phase 
is the two-magnon peak.  
The $B_{\rm 2g}$ scattering intensity in the $(ab)$ spectra is weak.  
If only the nearest neighbor exchange interaction is taken into account, 
the two-magnon peak energy is $2.76J$ $(3.38J^*)$ \cite{Canali}, where 
$J^*$ is the exchange interaction energy with the correction of the quantum 
spin effect of $S=1/2$ by Oguchi.  
The obtained $J^*$ is 949 cm$^{-1}$ (118 meV).  
The maximum energy in the density of independent two magnon states is 
$4J$ $(4.63J^*)$ \cite{Canali}.  
The energies of the experimentally obtained two-magnon Raman peak and the independent 
two magnon peak are shown by the vertical dashed lines.  
The density of independent two magnon states calculated using single $J^*$ is 
shown by the solid line in the upper panel of Fig. 11(b).  
Coldea {\it et al}. \cite{Coldea} obtained the full spin wave dispersion in 
neutron scattering.  
The fitted parameters are $J^*=104$ and $J^*_2=-18$ meV, where $J^*_2$ is 
the second nearest neighbor exchange interaction energy.  
The density of independent two magnon states is shown by the dashed line.  
The two-magnon scattering peak and the independent two magnon peak widely 
separates in the cuprates compared to BaFe$_2$As$_2$, because the spins in 
cuprates are $S=1/2$ and those in iron pnictide are $S=1$.  
There is small spectral weight above the independent two magnon peak energy 
in the insulating phase, indicating that the higher order multi-magnon 
scattering intensity is small \cite{Singh}.  
The two-magnon peak remains as temperature increases above 
the AF transition temperature $T_{\rm N}$.  
In hole-doped La$_{2-x}$Sr$_x$CuO$_4$ the peak energy shifts to low 
energy with broadening.  
The magnetic scattering cannot be derived from the same process as in the 
insulating phase, because the integrated intensity does not decrease.  
If the decrease of the peak energy is caused by the shortened correlation 
length, the integrated intensity decreases \cite{Tohyama}.  

In Nd$_2$CuO$_4$ of Fig. 11(c) the two-magnon spectra in the insulating 
phase are nearly the same as La$_2$CuO$_4$.  
The peak energy is 2890 cm$^{-1}$ and the obtained $J^*$ is 855 cm$^{-1}$
The density of independent two magnon states using only $J^*$ is shown by 
the solid curve in the upper panel in Fig. 11(c).  
The electron doping causes the abrupt shift of the peak energy to 
high energy at the insulator-metal transition.  
The peak energy little shifts as the electron density increases in the metallic phase. 
The high-energy shift in the metallic phase cannot be interpreted by the 
Coulomb interaction induced two-magnon scattering mechanism.  
The spectral weights do not decrease in the metallic phase similarly to 
La$_{2-x}$Sr$_x$CuO$_4$.  
The $B_{\rm 2g}$ spectral intensity observed in the $(ab)$ spectra 
increases to nearly the same magnitude of the $B_{\rm 1g}$ intensity in 
the $(xy)$ spectra as carrier density increases.  
The magnetic scattering spectra in the metallic phase may be interpreted 
by electronic scattering with the self-energy part of the magnetic 
component as discussed in the following.  

The electronic state of a hole moving in the AF lattice is extensively 
investigated in cuprates.  
Hopping of a hole is caused by the back-hopping of an electron with the opposite 
spin from the nearest neighbor site.  
Therefore the moving hole leaves behind an overturned spin trace in the 
AF spin lattice. 
The increased magnetic energy prevents the hole hopping.
A pair of opposite spins on the trace may return to the original directions 
by the quantum spin fluctuation, one and a half rotation of a hole around the 
nearest-neighbor four Cu square sites \cite{Trugman}, the short range 
AF correlation, and the hopping of other holes.  
The electronic state was calculated, for example, by the string model taking 
into account the quantum spin 
fluctuation \cite{Brinkman,Shraiman,Marsiglio,Dagotto,Martinez,Liu2,Lee,Manousakis}.  
The hole sees a linearly rising potential on increasing the migration distance 
in the limit of $t>>J$, which makes a series of discrete energy levels 
\begin{eqnarray}
E_n/t=-b+a_n(J/t)^{2/3},
\end{eqnarray}
where $a_n$ are the eigen values of a dimensionless Airy equation, 
$a_1=2.16$, $a_2=5.46$, $a_3=7.81$, and $b=3.28$ at $(\pi/2, \pi/2)$ \cite{Liu2}. 
The discrete peak height decreases with increasing $n$ and connects to a 
continuum at high energies.  
The ground-state $E_1$ is a delta-function and the width of the second peak is 
about $2J$ at $J=0.1t$.  
The energy difference between $E_1$ and $E_2$ is 4700 cm$^{-1}$ using the widely 
accepted values of $t=0.4$ eV and $J/t=0.3$ \cite{Kim}.  
This energy is close to the broad peak energy in Nd$_{2-x}$Ce$_x$CuO$_4$ with 
$x=0.14$.  
The discrete levels cannot be detected in Raman scattering, because the levels 
have dispersions.  
The nearly constant peak energy as the carrier density increases in the 
metallic phase of Nd$_{2-x}$Ce$_x$CuO$_4$ is successfully explained by the 
string model.  

In the metallic phase of the hole doped cuprate La$_{2-x}$Sr$_x$CuO$_4$, the 
magnetic peak energy continuously decreases with increasing the carrier density.  
The mean free path in the underdoped phase of the hole doped cuprate is one order 
shorter than that of electron doped cuprate.  
The migration distance is not as long as to allow the linearly rising 
potential approximation.  
The AF correlation length decreases with the increase of the hole 
density \cite{Birgeneau}.
The electron Green function has a self-energy part of excited spins by the 
hole hopping.  
The electron spectral functions presented by the imaginary part of the retarded 
Green function has the coherent component near the original electronic energy 
and the incoherent component near the spin excitation energies.  
The $\Delta k\approx 0$ transition between the electron spectral functions 
gives the spectral weight at the spin excitation energies.  
It is in contrast to the uncorrelated electron spectral function which is 
presented by the delta function at the energy of the dispersion and the 
$\Delta k\approx 0$ transition gives only $\Delta \epsilon \approx 0$ excitations.  

The very broad two-magnon peak in BaFe$_2$As$_2$ is compared to the magnetic 
excitation peak in the metallic cuprate.  
The hole density is 0.081 and the electron density is 0.069 per iron atom at 
140 K in BaFe$_2$As$_2$ \cite{YiSDW}.  
The spectra little changes by electron doping in BaFe$_{2-x}$Co$_x$As$_2$.  
The large spectral weight above the maximum of the independent 
two-magnon energy and the nearly independent peak energy on the electron 
density are similar to Nd$_{2-x}$Ce$_x$CuO$_4$.  
However, the peak energy is given by the two-magnon excitation energy estimated 
from the spin wave dispersion in BaFe$_2$As$_2$, in contrast to the much higher 
energy in Nd$_{2-x}$Ce$_x$CuO$_4$.

\subsection{Particle-hole excitations in the itinerant model}
In the itinerant spin model the particle-hole continuum is created 
above the spin wave dispersion \cite{Kariyado,Brydon,Kaneshita,Knolle}.  
The spin wave dispersion strongly decays when the energy overlaps with the 
particle-hole continuum.  
Kaneshita and Tohyama \cite{Kaneshita} calculated the imaginary part of the 
susceptibility using the five-band model.  
The spin wave dispersion becomes clear as the magnetic moment increases.  
The full spin wave dispersion with the maximum energy 0.25 eV and the 
dispersion with the maximum energy 0.4 eV above it are obtained at the 
magnetic moment $M=0.8\mu_B/{\rm Fe}$.  
The calculation showed that the integrated intensity over $k$ increases 
from energy zero to 0.2 eV.  
The intensity is larger in the paramagnetic phase than in the 
AF phase.  
In the Raman scattering two particle-two hole excitations are observed.  
The scattering intensity decreases on increasing temperature.
The increase of intensity in the paramagnetic phase is not observed.  
The calculation did not show whether the intensity maximum occurs at twice 
the top of the spin wave dispersion or not.  
The imaginary part of the susceptibility for the longitudinal spin excitations 
and the charge excitations has large intensity from 0.2 to 0.4 eV.  
The two particle-two hole excitations may observe from 0.4 to 0.8 eV.  
The observed spectra have the broad peak as shown in Fig. 11(a).

\section{Conclusion}
The change of the electronic and magnetic states at the SDW transition in 
BaFe$_2$As$_2$ was investigated by Raman scattering.  
The tight binding band model with two orbitals disclosed that the electronic excitations near 
the Dirac node and the anti-node in the SDW state have different symmetries.  
Utilizing this selection rule the electronic excitations near the Dirac node 
and the anti-node were separately obtained.  
The electronic excitations near the anti-node have critical fluctuation above 
$T_{\rm SDW}$ and change into the gap structure by the first order transition, 
while those near the Dirac node gradually changes.  
As for the magnetic scattering the selection rule of two-magnon scattering in 
the stripe spin structure was obtained.  
The magnetic exchange interaction energies are not presented by the superexchange 
interaction model, but the second derivative of the total energy of the 
long-range stripe spin structure with respect to the moment directions.  
The magnetic excitation peak survives far above $T_{\rm SDW}$ and in the 
superconductor phase of BaFe$_{2-x}$Co$_x$As$_2$, indicating that the short-range 
spin correlation remains in the two-dimensional magnet.  
The two-magnon peak energy is well understood by the conventional two-magnon 
scattering process in insulator.  
However, the high-energy spectral weight above twice the maximum energy of the 
spin wave may be explained by the string model in the 
localized spin picture or the particle-hole excitations in the itinerant picture.  
The difference from the high temperature superconducting cuprates were discussed. 

\begin{acknowledgment}
This work was supported by E Research Project on Iron Pnictides 
(TRIP), Japan Science and Technology Agency (JST).

\end{acknowledgment}


\begin{thebibliography}{9}
\bibitem{Kamihara}Y. Kamihara, T. Watanabe, M. Hirano, and H. Hosono, 
J. Am. Chem. Soc. {\bf 130} (2008) 3296.

\bibitem{Ren}Z.-A. Ren, G.-C. Che, X.-L. Dong, J. Yang, W. Lu, W. Yi, X.-L. Shen, 
Z.-C. Li, L.-L. Sun, F. Zhou, and Z.-X. Zhao: Europhys. Lett. {\bf 83} (2008) 17002.

\bibitem{Chen}G. F. Chen, Z. Li, D. Wu, G. Li, W. Z. Hu, J. Dong, P. Zheng, 
J. L. Luo, and N. L. Wang: Phys. Rev. Lett. {\bf 100} (2008) 247002.

\bibitem{Rotter}M. Rotter, M. Tegel, .D. Johrendt, I. Schellenberg, W. Hermes, 
and R. P\"ottgen: Phys. Rev. B {\bf 78} (2008) 020503(R).

\bibitem{Rotter2}M. Rotter, M. Tegel, and D. Johrendt: Phys. Rev. Lett. {\bf 101} (2008) 
107006.

\bibitem{Torikachvili}M. S. Torikachvili, S. L. Bud'ko, N. Ni, and 
P. C. Canfield: Phys Rev. Lett. {\bf 101} (2008) 057006.

\bibitem{Thio}T. Thio, T. R. Thurston，N. W. Preyer, P. J. Picone, M. A. Kastner, 
H. P. Jenssen, D. R. Gabbe, C. Y. Chen, R. J. Birgeneau, and A. Aharony: 
Phys. Rev. {\bf 38} (1988) 905.

\bibitem{Ishikado}M. Ishikado, R. Kajimoto, S. Shamoto, M. Arai, 
A. Iyo, K. Miyazawa, P. M. Shirage, H. Kito,
H. Eisaki, S. Kim, H. Hosono, T. Guidi,
R. Bewley, and S. M. Bennington: J. Phys. Soc. Japan, {\bf 78} (2009) 043705.

\bibitem{Diallo2010}S. O. Diallo, D. K. Pratt, R. M. Fernandes, W. Tian, 
J. L. Zarestky, M. Lumsden, T. G. Perring, C. L. Broholm, N. Ni, S. L. Bud'ko, 
P. C. Canfield, H.-F. Li, D. Vaknin, A. Kreyssig, A. I. Goldman, and R. J. McQueeney: 
Phys. Rev. Rev. B {\bf 81} (2010) 214407.

\bibitem{Matan}K. Matan, R. Morinaga, K. Iida, and T. J. Sato: 
Phys. Rev. B {\bf 79} (2009) 054526.  

\bibitem{Sugai}S. Sugai, Y. Mizuno, K. Kiho, M. Nakajima, C. H. Lee, A. Iyo, 
H. Eisaki, and S. Uchida: Phys. Rev. B {\bf 82} (2010) 140504(R).

\bibitem{Okazaki}K. Okazaki, S. Sugai, S. Niitaka, and H. Takagi: Phys. Rev. 
B {\bf 83} (2011) 035103.

\bibitem{Kaneko}K. Kaneko, A. Hoser, N. Caroca-Canales, A. Jesche, C. Krellner, 
O. Stockert, and C. Geibel: Phys. Rev. B {\bf 78} (2008) 212502.

\bibitem{Ewings}R. A. Ewings, T. G. Perring, R. I. Bewley, T. Guidi, 
M. J. Pitcher, D. R. Parker, S. J. Clarke, and A. T. Boothroyd: Phys. Rev. B {\bf 78} (2008) 220501(R).

\bibitem{Goldman}A. I. Goldman, D. N. Argyriou, B. Ouladdiaf, T. Chatterji, 
A. Kreyssig, S. Nandi, N. Ni, S. L. Bud'ko, P. C. Canfield, and R. J. McQueeney: 
Phys. Rev. B {\bf 78} (2008) 100506(R).

\bibitem{Johannes}M. D. Johannes and I. I. Mazin: Phys. Rev. B {\bf 79} (2009) 220510(R).

\bibitem{Huang}Q. Huang, Y. Qiu,1,2 Wei Bao, M. A. Green, J.W. Lynn, 
Y. C. Gasparovic, T. Wu, G. Wu, and X. H. Chen: Phys. Rev. Lett. {\bf 101} (2008) 257003.

\bibitem{Diallo2009}S. O. Diallo, V. P. Antropov, T. G. Perring, C. Broholm, 
J. J. Pulikkotil, N. Ni, S. L. Bud'ko, P. C. Canfield, A. Kreyssig, A. I. Goldman, 
and R. J. McQueeney: Phys. Rev. Lett. {\bf 102} (2009) 187206.

\bibitem{Mazin}I. I. Mazin, D. J. Singh, M. D. Johannes, and M. H. Du: 
Phys. Rev. Lett. {\bf 101} (2008) 057003.

\bibitem{Kuroki}K. Kuroki, S. Onari, R. Arita, H. Usui, Y. Tanaka, H. Kontani, 
and H. Aoki: Phys. Rev. Lett. {\bf 101} (2008) 087004.

\bibitem{Dong}J. Dong, H. J. Zhang, G. Xu, Z. Li, G. Li, W. Z. Hu, D. Wu, 
G. F. Chen, X. Dai, J. L. Luo, Z. Fang, and N. L. Wang: Europhys. 
Lett. {\bf 83} (2008) 27006.  

\bibitem{Ding}H. Ding, P. Richard, K. Nakayama, K. Sugawara, T. Arakane, 
Y. Sekiba, A. Takayama, S. Souma, T. Sato, T. Takahashi, Z. Wang, X. Dai, 
Z. Fang, G. F. Chen, J. L. Luo, and N. L. Wang: Europhys. Lett. {\bf 83} (2008) 47001.

\bibitem{Terashima}K. Terashima, Y. Sekiba, J. H. Bowen, K. Nakayama, T. Kawahara, 
T. Sato, P. Richard, Y.-M. Xu, L. J. Li, 
G. H. Cao, Z.-A. Xu, H. Ding, and T. Takahashi: Proceedings of the National 
Academy of Sciences of the United States of America, {\bf 106},  No. 18 (2009) 
7330.

\bibitem{Li}S. Li, C. de la Cruz, Q. Huang, Y. Chen, J. W. Lynn, J. Hu, 
Y.-L. Huang, F.-C. Hsu, 
K.-W. Yeh, M.-K. Wu, and P. Dai: Phys. Rev. B {\bf 79} (2009) 054503.

\bibitem{Bao}W. Bao, Y. Qiu, Q. Huang, M. A. Green, P. Zajdel, M. R. Fitzsimmons, 
M. Zhernenkov, S. Chang, M. Fang, B. Qian, E. K. Vehstedt, J. Yang, H. M. Pham, 
L. Spinu, and Z. Q. Mao: Phys. Rev. Lett. {\bf 102} (2009) 247001.

\bibitem{Wang}F. Wang, H. Zhai, Y. Ran, A. Vishwanath, and D.-H. Lee: arXis:0805.3343.

\bibitem{Ran}Y. Ran, F. Wang, H. Zhai, A. Vishwanath, and D.-H. Lee: 
Phys. Rev. B {\bf 79} (2009) 014505.

\bibitem{Harrison}N. Harrison and S. E. Sebastian: Phys. Rev. B {\bf 80} (2009) 224512.

\bibitem{Morinari}T. Morinari, E. Kaneshita, and T. Tohyama: Phys. Rev. 
Lett. {\bf 105} (2010) 037203.

\bibitem{Zhao}J. Zhao, D.-X. Yao, S. Li, T. Hong, Y. Chen, S. Chang, 
W. Ratcliff II, J.W. Lynn, H. A. Mook, G. F. Chen, J. L. Luo, N. L. Wang, 
E.W. Carlson, J. Hu, and P. Dai: Phys. Rev. Lett. {\bf 101} (2008) 167203.

\bibitem{Yildirim}T. Yildirim: Phys. Rev. Lett. {\bf 101} (2008) 057010,

\bibitem{Yin}Z. P. Yin, S. Leb\`egue, M. J. Han, B. P. Neal, S.Y. Savrasov, 
and W. E. Pickett: Phys. Rev. Lett. {\bf 101} (2008) 047001,

\bibitem{Si}Q. Si and E. Abrahams:  Phys. Rev. Lett. {\bf 101} (2008) 076401,

\bibitem{Han}M. J. Han, Q. Yin, W. E. Pickett, and S. Y. Savrasov: Phys. 
Rev. Lett. {\bf 102} (2009) 107003.

\bibitem{Dong2}J K Dong, L Ding, H Wang, X F Wang, T Wu, G Wu2, 
X H Chen and S Y Li: New J. Phys. {\bf 10} (2008) 123031.

\bibitem{Chu2009}J.-H. Chu, J. G. Analytis, C. Kucharczyk, and I. R. Fisher: 
Phys. Rev. B {\bf 79} (2009) 014506.

\bibitem{Kitagawa}K. Kitagawa, N. Katayama, K. Ohgushi, and M. Takigawa: 
J. Phys. Soc. Japan, {\bf 78} (2009) 063706.

\bibitem{Baek}S.-H. Baek, N. J. Curro, T. Klimczuk, E. D. Bauer, F. Ronning, 
and J. D. Thompson: Phys. Rev. B {\bf 79} (2009) 052504.

\bibitem{Tanatar}M. A. Tanatar, N. Ni, G. D. Samolyuk, S. L. Bud'ko, 
P. C. Canfield, and R. Prozorov: Phys. Rev. B {\bf 79} (2009) 134528.

\bibitem{Cruz}C. de la Cruz, Q. Huang, J. W. Lynn, J. Li, W. Ratcliff II, 
J. L. Zarestky, H. A. Mook, G. F. Chen, J. L. Luo, N. L. Wang, and P. Dai: Nature, {\bf 453} (2008) 899.

\bibitem{McQueeney} R. J. McQueeney, S. O. Diallo, V. P. Antropov, 
G. D. Samolyuk, C. Broholm, N. Ni, S. Nandi, M. Yethiraj, J. L. Zarestky, 
J. J. Pulikkotil, A. Kreyssig, M. D. Lumsden, B. N. Harmon, P. C. Canfield, 
and A. I. Goldman: Phys. Rev. Lett. {\bf 101} (2008) 
227205.

\bibitem{Qi}X.-L. Qi, S. Raghu, C.-X. Liu, D. J. Scalapino and S.-C. Zhang: 
arXiv:0804.4332.

\bibitem{Raghu}S. Raghu, X.-L. Qi, C.-X. Liu, D. J. Scalapino, and S.-C. Zhang: 
Phys. Rev. B {\bf 77} (2008) 220503(R).

\bibitem{Graser}S. Graser, T. A. Maier, P. J. Hirschfeld, and 
D. J. Scalapino: New J. Phys. {\bf 11} (2009) 025016.

\bibitem{Andersen}O. K. Andersen and L. Boeri: Ann. Phys. (Berlin) {\bf 523} (2011) 8.

\bibitem{Eremin}I. Eremin and A. V. Chubukov: Phys. Rev. B {\bf 81} (2010) 024511.

\bibitem{Richard}P. Richard, K. Nakayama, T. Sato, M. Neupane, Y.-M. Xu, J. H. Bowen, 
G. F. Chen, J. L. Luo, N. L. Wang, X. Dai, Z. Fang, H. Ding, and T. Takahashi: 
Phys. Rev. Lett. {\bf 104} (2010) 137001.

\bibitem{Novoselov}K. S. Novoselov, A. K. Geim, S. V. Morozov, D. Jiang, 
M. I. Katsnelson, I. V. Grigorieva, S. V. Dubonos, and A. A. Firsov: Nature, 
{\bf 438} (2005) 197.

\bibitem{SugaiSDWgap}S. Sugai, Y. Mizuno, R. Watanabe, T. Kawaguchi, K. Takenaka, 
H. Ikuta, Y. Takayanagi, N. Hayamizu, and Y. Sone: arXiv:1010.6151.

\bibitem{Chauviere2}L. Chauvi\`ere, Y. Gallais, M. Cazayous, M. A. M\'easson, 
A. Sacuto, D. Colson, and A. Forget: Phys. Rev. B {\bf 82} (2010) 180521(R).

\bibitem{Hu}W. Z. Hu, J. Dong, G. Li, Z. Li, P. Zheng, G. F. Chen, J. L. Luo, 
and N. L. Wang: Phys. Rev. Lett. {\bf 101} (2008) 257005.

\bibitem{Yang}L. X. Yang, Y. Zhang, H. W. Ou, J. F. Zhao, D. W. Shen, B. Zhou, 
J. Wei, F. Chen, M. Xu, C. He, Y. Chen, Z. D. Wang, X. F. Wang, T. Wu, G. Wu, 
X. H. Chen, M. Arita, K. Shimada, M. Taniguchi, Z. Y. Lu, T. Xiang, and D. L. Feng: 
Phys. Rev. Lett. {\bf 102} (2009) 107002.

\bibitem{Zhang}Y. Zhang, J. Wei, H. W. Ou, J. F. Zhao, B. Zhou, F. Chen, M. Xu, 
C. He, G. Wu, H. Chen, M. Arita, K. Shimada, H. Namatame, M. Taniguchi, X. H. Chen, 
and D. L. Feng: Phys. Rev. Lett. {\bf 102} (2009) 127003,

\bibitem{Liu}C. Liu, T. Kondo, N. Ni, A. D. Palczewski, A. Bostwick, G. D. Samolyuk, 
R. Khasanov, M. Shi, E. Rotenberg, S. L. Bud'ko, P. C. Canfield, and A. Kaminski: 
Phys. Rev. Lett. {\bf 102} (2009) 167004.

\bibitem{Fink}J. Fink, S. Thirupathaiah, R. Ovsyannikov, H. A. D\"urr, R. Follath, 
Y. Huang, S. de Jong, M. S. Golden, Y.-Z. Zhang, H. O. Jeschke, R. Valent\'i, 
C. Felser, S. D. Farahani, M. Rotter, and D. Johrendt: Phys. Rev. B {\bf 79} 
(2009) 155118.

\bibitem{YiSDW}M. Yi, D. H. Lu, J. G. Analytis, J.-H. Chu, S.-K. Mo, R.-H. He, 
M. Hashimoto, R. G. Moore, I. I. Mazin, D. J. Singh, Z. Hussain, I. R. Fisher, 
and Z.-X. Shen: Phys. Rev. B {\bf 80} (2009) 174510.

\bibitem{Yi}M. Yi, D. H. Lu, J. G. Analytis, J.-H. Chu, S.-K. Mo, R.-H. He, 
R. G. Moore, X. J. Zhou, G. F. Chen, J. L. Luo, N. L. Wang, Z. Hussain, 
D. J. Singh, I. R. Fisher, and Z.-X. Shen: Phys. Rev. B {\bf 80} (2009) 024515.

\bibitem{Lester}C. Lester, J.-H. Chu, J. G. Analytis, T. G. Perring, I. R. Fisher, 
and S. M. Hayden: Phys. Rev. Rev. B {\bf 81} (2010) 064505.

\bibitem{Zhao2}J. Zhao, D. T. Adroja, D.-X. Yao, R. Bewley, S. Li, X. F.Wang, 
G.Wu, X. H. Chen, J. Hu, and P. Dai: Nature Phys. {\bf 5} (2009) 555.

\bibitem{Kariyado}T. Kariyado and M. Ogata: J. Phys. Soc. Japan, {\bf 78} (2009) 043708.

\bibitem{Brydon}P. M. R. Brydon and C. Timm: Phys. Rev. B {\bf 80} (2009) 174401.

\bibitem{Kaneshita}E. Kaneshita and T. Tohyama: Phys. Rev. B {\bf 82} (2010) 094441.

\bibitem{Knolle}J. Knolle, I. Eremin, A. V. Chubukov, and R. Moessner: Phys. Rev. B 
{\bf 81} (2010) 140506(R).

\bibitem{Brinkman}W. F. Brinkman and T. M. Rice: Phys. Rev. B {\bf 2} (1970) 1324.

\bibitem{Shraiman}B. I. Shraiman and E. D. Siggia: Phys. Rev. Lett. {\bf 61} (1988) 467.

\bibitem{Marsiglio}F. Marsiglio, A. E. Ruckenstein, S. Schmitt-Rink, and 
C. M. Varma: Phys. Rev. B {\bf 43} (1991) 10882.

\bibitem{Dagotto}E. Dagotto, R. Joynt, A. Moreo, S. Bacci, and E. Gagliano: 
Phys. Rev. B {\bf 41} (1990) 9049.

\bibitem{Martinez}G. Martinez and P. Horsch: Phys. Rev. B {\bf 44} (1991) 317.

\bibitem{Liu2}Z. Liu and E. Manousakis: Phys. Rev. B {\bf 45} (1992) 2425.

\bibitem{Lee}P. A. Lee, N. Nagaosa, and X.-G. Wen: Rev. Mod. Phys. {\bf 78} (2006) 17.

\bibitem{Manousakis}E. Manousakis: Phys. Rev. B {\bf 75} (2007) 035106.

\bibitem{Wolff}P. A. Wolff: Phys. Rev. Lett., {\bf 16} (1966) 225.

\bibitem{Platzman}P. M. Platzman and N. Tzoar: Phys. Rev. {\bf 182} (1969) 510.

\bibitem{Klein}M. V. Klein and S. B. Dierker: Phys. Rev. B {\bf 29} (1984) 4976.

\bibitem{Cardona}M. Cardona: Physica C {\bf 317-318} (1999) 30.

\bibitem{MazinGap}I. I. Mazin, T. P. Devereaux, J. G. Analytis, J.-H. Chu, 
I. R. Fisher, B. Muschler, and R. Hackl: Phys. Rev. B {\bf 82} (2010) 180502(R).

\bibitem{Muschler}B. Muschler, W. Prestel, R. Hackl, T. P. Devereaux, 
J. G. Analytis, J.-H. Chu, and I. R. Fisher: Phys. Rev. B {\bf 80} (2009) 180510(R).

\bibitem{Moreo}A. Moreo, M. Daghofer, J. A. Riera, and E. Dagotto: Phys. Rev. B 
{\bf 79} (2009) 134502.

\bibitem{Zhang2}J. Zhang, R. Sknepnek, R. M. Fernandes, and J. Schmalian: Phys. 
Rev. B {\bf 79} (2009) 220502(R).

\bibitem{Litvinchuk}A. P. Litvinchuk, V. G. Hadjiev, M. N. Iliev, Bing Lv, 
A. M. Guloy, and C. W. Chu: Phys. Rev. B {\bf 78} (2008) 
060503(R).

\bibitem{Rahlenbeck}M. Rahlenbeck, G. L. Sun, D. L. Sun, C. T. Lin, B. Keimer, 
and C. Ulrich: Phys. Rev. B {\bf 80} (2009) 064509. 

\bibitem{Choi}K.-Y. Choi, D. Wulferding, P. Lemmens, N. Ni, S. L. Bud'ko, 
and P. C. Canfield: Phys. Rev. B {\bf 78} (2008) 212503.

\bibitem{Chauviere}L. Chauvi\`ere, Y. Gallais, M. Cazayous, A. Sacuto, and 
M. A. M\'easson: Phys. Rev. B {\bf 80} (2009) 094504.

\bibitem{KaneshitaPRL}E. Kaneshita, T. Morinari, and T. Tohyama: Phys. Rev. 
Lett. {\bf 103} (2009) 247202.

\bibitem{Graser2010}S. Graser, A. F. Kemper, T. A. Maier, H.-P. Cheng, 
P. J. Hirschfeld, and D. J. Scalapino: Phys. Rev. B {\bf 81} (2010) 214503.

\bibitem{Mahan}G. D. Mahan: {\it Many-Particle Physics}, 
(Kluwer Academic/Plenum Publishers, New York, 2000).  

\bibitem{Cao}C. Cao, P. J. Hirschfeld, and H.-P. Cheng: Phys. Rev. B {\bf 77} (2008) 
220506(R).

\bibitem{Ikeda}H. Ikeda, R. Arita, and J. Kune\v{s}: Phys. Rev. B {\bf 81} (2010) 054502.

\bibitem{Kuroki2}K. Kuroki, H. Usui, S. Onari, R. Arita, and H. Aoki: 
Phys. Rev. B {\bf 79} (2009) 224511.

\bibitem{Akrap}A. Akrap, J. J. Tu, L. J. Li, G. H. Cao, Z. A. Xu, and C. C. Homes: 
Phys. Rev. B {\bf 80} (2009) 180502(R).

\bibitem{Ni}N. Ni, M. E. Tillman, J.-Q. Yan, A. Kracher, S. T. Hannahs, 
S. L. Bud'ko, and P. C. Canfield: Phys. Rev. B {\bf 78} (2008) 214515.

\bibitem{Drew}A. J. Drew, Ch. Niedermayer, P. J. Baker, F. L. Pratt, S. J. Blundell, 
T. Lancaster, R. H. Liu, G.Wu, X. H. Chen, I.Watanabe, V. K. Malik, A. Dubroka, 
M. R\"ossle, K.W. Kim, C. Baines, and C. Bernhard: Nature Materials, {\bf 8} (2009) 310.

\bibitem{Fernandes}R. M. Fernandes, D. K. Pratt, W. Tian, J. Zarestky, A. Kreyssig, 
S. Nandi, M. G. Kim, A. Thaler, N. Ni, P. C. Canfield, R. J. McQueeney, J. Schmalian, 
and A. I. Goldman: Phys. Rev. B {\bf 81} (2010) 140501(R).

\bibitem{Moriya}T. Moriya: J. Appl. Phys. {\bf 39} (1968) 1042.

\bibitem{Fleury}P. A. Fleury, S. P. S. Porto, and R. Loudon: Phys. Rev. 
Lett. {\bf 18} (1967) 658.

\bibitem{Fleury1968}P. A. Fleury: Phys. Rev. Lett. {\bf 21} (1968) 151.

\bibitem{Elliott}R. J. Elliott, M. F. Thorpe, G. F. Imbusch, R. Loudon, 
and J. B. Parkinson: Phys. Rev. Lett. {\bf 21} (1968) 147.

\bibitem{Elliott2}R. J. Elliott, M. F. Thorpe: J. Phys. C (Solid St. Phys.), 
Ser 2, {\bf 2} (1969) 1630.

\bibitem{Parkinson}J. B. Parkinson: J. Phys. C (Solid St. Phys.), Ser 2, 
{\bf 2} (1969) 2012.

\bibitem{Chentwomag}C.-C. Chen, C. J. Jia, A. F. Kemper, R. R. P. Singh, and T. P. Devereaux: 
Phys. Rev. Lett. {\bf 106} (2011) 067002.

\bibitem{Singh}R. R. P. Singh，P. A. Fleury，K. B. Lyons，and P. E. Sulewski: 
Phys. Rev. Lett. {\bf 62} (1989) 2736.

\bibitem{Canali}C. M. Canali and S. M. Girvin: Phys. Rev. B {\bf 45} (1992) 7127.

\bibitem{Coldea}R. Coldea, S. M. Hayden, G. Aeppli, T. G. Perring, C. D. Frost, 
T. E. Mason, S.-W. Cheong, and Z. Fisk, Phys. Rev. Lett. {\bf 86} 
(2001) 5377.  

\bibitem{Tohyama}T. Tohyama: Phys. Rev. B {\bf 70} (2004) 174517.

\bibitem{Trugman}S. A. Trugman: Phys. Rev. B {\bf 37} (1988) 1597.

\bibitem{Kim}C. Kim, P. J. White, Z.-X. Shen, T. Tohyama, Y. Shibata, S. Maekawa, 
B. O. Wells, Y. J. Kim, R. J. Birgeneau, and M. A. Kastner: Phys. Rev. Lett. 
{\bf 80} (1998) 4245.

\bibitem{Birgeneau}R. J. Birgeneau，D. R. Gabbe，H . P. Jenssen，M . A. Kastner，
P. J. Picone，T. R. Thurston, G. Shirane，Y. Endoh，M. Sato，K. Yamada, 
Y. Hidaka，M . Oda，Y. Enomoto，M . Suzuki，and T. Murakami: Phys. Rev. B {\bf 38} 
(1988) 6614.  

\end{thebibliography}
\end{document}